\documentclass{jfm}
\usepackage{graphicx}
\usepackage{epstopdf, epsfig}
\usepackage{mathrsfs}
\pdfoutput=1
 \usepackage[final]{changes} 

\newcommand{\pardt}[1]{\frac{\partial #1}{\partial t}}

\newcommand{\pardX}[1]{\frac{\partial #1}{\partial X}}
\newcommand{\pardY}[1]{\frac{\partial #1}{\partial Y}}
\newcommand{\pardZ}[1]{\frac{\partial #1}{\partial Z}}

\newcommand{\pardxx}[1]{\frac{\partial^2 #1}{\partial x^2}}
\newcommand{\pardyz}[1]{\frac{\partial^2 #1}{\partial y \partial z}}
\newcommand{\pardyy}[1]{\frac{\partial^2 #1}{\partial y^2}}
\newcommand{\pardzz}[1]{\frac{\partial^2 #1}{\partial z^2}}

\newcommand\Frou{\mbox{\textit{Fr}}} 
\newcommand\Sc{\mbox{\textit{Sc}}}   
\newcommand\Ri{\mbox{\textit{Ri}}}   

\usepackage{floatrow}
 \usepackage[caption=false,labelformat=simple]{subfig}

\shorttitle{Instability of tilted stratified viscous shear flow}
\shortauthor{L. Fung and Y. Hwang}

\title{Instability of tilted shear flow in a strongly stratified and viscous medium}
\author{Lloyd Fung\aff{1}
  \corresp{\email{lloyd.fung@imperial.ac.uk}},
 \and Yongyun Hwang\aff{1}}

\affiliation{\aff{1}Department of Aeronautics, Imperial College London, South Kensington, SW7 2AZ, London, UK}

\begin{document}
\maketitle
\begin{abstract}
A linear stability analysis is performed on a tilted parallel wake in a strongly stratified fluid at low Reynolds numbers. A particular emphasis of the present study is given to the understanding of the low-Froude-number mode observed by the recent experiment (Meunier, \textit{J. Fluid
Mech.}, vol. 699, 2012, pp. 174-197). In the limit of low Froude number, the linearised equations of motion can be reduced to the Orr-Sommerfeld equation on the horizontal plane, except the viscous term that contains vertical dissipation. Based on this equation, it is proposed that the low-Froude-number mode would be a horizontal inflectional instability, and should remain two dimensional at small tilting angles. To support this claim, the asymptotic regime where this equation would be strictly valid is subsequently discussed in relation to previous arguments on the proper vertical length scale. Furthermore, the absolute and convective instability analysis of parallel wake is performed, revealing qualitatively good agreement with the experimental result. The low-Froude-number mode is found to be stabilised on increasing Froude number, as is in the experiment. The emergence of small vertical velocity at finite Froude number, the size of which is proportional to the square of Froude number, plays the key role in the stabilisation. It modifies the inflectional instability and is also associated with the paradoxically stabilising buoyancy on increasing Froude number. Lastly, we proposed some possible behaviours of the base flow when the tilting angle changes, and they may provide a better approximation to produce the behaviour consistent with the experiment. 
\end{abstract}

\section{Introduction\label{sec:Introduction}}

\subsection{\added{Shear flow instability in strongly stratified and viscous medium}}
The wake behind a bluff body has long been studied as one of the fundamental open shear flows in hydrodynamic stability, but few have studied the flow in a stratified medium at low Reynolds numbers.
Much of the earlier research focused on the stabilising effect of stratification. Many have demonstrated that stratification has a stabilising effect on shear flows, theoretically \citep{Koppel1964}, numerically \citep{Gage1968,Hazel1972} and experimentally  \citep{Boyer1989}. All of the results confirmed \replaced{Howard's theorem \citep{Howard1961,Miles1961}, which stated }{the seminal finding by \mbox{\citet{Howard1961}} and \mbox{\citet{Miles1961}}, who demonstrated }that the stability property of a vertical shear flow under stable linear stratification is governed by the `local' Richardson number:
\begin{equation}
\Ri=N^{*2}\left(\frac{dU^{*}}{dZ^{*}}\right)^{-2},\label{eq:Howards_criterion}
\end{equation}
where the superscript $^*$ indicates dimensional quantities, $N^{*}$ is the Brunt-V\"{a}is\"{a}l\"{a} frequency, $U^*$ the base-flow velocity and $dU^*/dZ^*$ the shear rate of the base-flow with $Z^*$ being the vertical direction. \added{While a criterion on $\Ri$ for stability had already been conjectured by \citet{Richardson1926}, \citet{Prandtl1930} and \citet{Taylor1931}, it was} \citet{Howard1961} \added{who elegantly }showed that \added{the flow becomes stable} if the local Richardson number everywhere is greater than or equal to $1/4$\deleted{, the flow becomes stable}. 
The implication of (\ref{eq:Howards_criterion}) is that the buoyancy force would stabilise the flow, while the shear would destabilise it, such that there exists a critical $\Ri$ that determines the necessary (but not sufficient) condition for flow instability. 

This criterion, however, applies only to the instabilities of vertical shear flows, as the strong stable stratification would inhibit vertical fluid motion. Indeed, when a shear flow exists in the horizontal plane, the stabilising mechanism is no longer at play.  \citet{Blumen1971,Blumen1975} first demonstrated the existence of horizontal shear instability by considering a base-flow profile that contains both horizontal and vertical shear. In the context of geophysical fluid dynamics, this horizontal shear instability has been known as barotropic instability  \citep[see ][]{PedloskyJoseph1987Gfd,Vallis2017}.
Recently, such a horizontal shear instability has also been confirmed even in strongly stratified flow where the base flow is not exactly aligned to the vertical direction. For example, both the stability analysis of a tilted stratified inviscid Bickley jet \citep{Candelier2011} and the experiment of a tilted stratified cylinder wake \citep{Meunier2012} have shown that there exists a horizontal shear instability arising at very low Froude number even for $\Ri>1/4$ everywhere.
In both of the investigations, the most unstable mode was shown to experience a branch switching behaviour as Froude number decreases (see figure 6 in \citet{Meunier2012} and figure 3 in \citet{Candelier2011}), indicating a new type of instability mode arising at low Froude number. 
With a further decrease of the Froude number, the low-Froude-number mode was found to be even more destabilised, and, interestingly, the mode structure appears to be quite similar to that of a typical inflectional instability. Furthermore, the experiment by \citet{Meunier2012} showed that the critical Reynolds number of the low-Froude-number mode returns to a value similar to that of a homogeneous wake (i.e. the high-Froude-number instability mode) as the Froude number reaches zero. 

\added{In the context of stratified turbulence, the horizontal shear has also been understood to play a significant role in turbulence production. \citet{Jacobitz1998} studied the effect of tilted uniform shear on turbulence under stable stratification. They found that the introduction of horizontal shear through tilting significantly increases turbulence production. \citet{Jacobitz2002} further compared turbulence statistics in horizontal uniform shear flow with those in vertical one under strong stratification. He found that the horizontal shear flow exhibits significantly larger turbulent velocity and density fluctuation than the vertical one. These results point to the important role of the horizontal shear in turbulence production in a strongly stratified flow, setting out the understanding of the genesis of turbulence from the horizontal shear.}

In the regime where the new instability mode arises in a titled shear flow (i.e. at low Froude number), the vertical length scale of the system has been shown to be determined by the interplay between the buoyancy force from the stratification and the fluid viscous force. In particular, \citet{Billant2001} proposed that, at high Reynolds numbers, the vertical length scale is proportional to the Froude number, while challenging the earlier argument of \citet{Lilly1983}, who proposed that highly stratified flow at high Reynolds number can be described only in terms of the two-dimensional dynamics on the horizontal plane. 
With this new scaling argument, the recent observations in the tilted stratified flows lead us to raise the following questions on the low Froude number instability mode:
\begin{enumerate}
\item If the Howard's rule does not apply to the horizontal instability mode, how can this mode be destabilised with increasing stratification, while maintaining the typical form of inflectional instability (i.e. von K\'arm\'an-vortex street in the experiment of \citet{Meunier2012})? 
\item As a result of the scaling proposed by \citet{Billant2001}, three-dimensional zigzag instability has been previously reported in strongly stratified flows \citep[see][]{Billant2000a,Billant2000b,Billant2000c}. Why was any structure, the vertical length scale of which is proportional to Froude number, not observed in the primary instability in the experiment by \citet{Meunier2012}? Is it possible to provide any theoretical justification that the low-Froude-number mode is inherently two dimensional in this particular case? In fact, this point may also be intricately linked to some `Squire-Yih-like' theorem for a flow configuration such as that of  \citet{Meunier2012}. 
\item While the experimental data of \citet{Meunier2012} suggests that the critical Reynolds number of the low-Froude-number instability mode is roughly independent of the tilting angle, the stability analysis of \citet{Candelier2011} showed that the growth rate of the instability is strongly dependent on the tilting angle. Here, we note that the analysis by \citet{Candelier2011} was carried out by prescribing a constant base flow while the tilting angle varies. However, the question of how the base flow is changed with respect to the tilting angle remains unanswered. This issue might be critical to address the difference between the experimental result of \citet{Meunier2012} and the theoretical one of \citet{Candelier2011}. 
\end{enumerate}

\subsection{\added{Contribution of the present study}}
\added{The objective of the present study is to gain better understanding on the low-Froude-number instability in titled shear flows under strong stratification by addressing the questions above. To this end, we perform a linear stability analysis of viscous parallel wake flow under strong stratification, especially focusing on the scaling and emergence of the low-Froude-number instability. Particular emphasis of the present study is given to address the following points:}
\begin{enumerate}
    \item \added{Derivation of the equation explicitly describing the horizontal shear-flow instability in the limit of low Froude number and low buoyancy Reynolds number;}
    \item \added{A Squire-like theorem for horizontal instability (i.e. barotropic instability) in a weakly titled shear flow at low buoynacy Reynolds numbers;}
    \item \added{Spatio-temporal stability analysis of tilted two-dimensional wake for qualitative comparison with the experimental data;}
    \item \added{The stabilisation mechanisms of horizontal shear flow instability with increase of Froude number;}
    \item \added{Identification of physical factors contributing to base-flow change in the experiment of \cite{Meunier2012} and the subsequent modelling of their roles. }
\end{enumerate} 
\added{It is important to mention that these points make the present studies clearly distinguished from previous investigations, such as \cite{Candelier2011} who studied tilted shear flow instability in the inviscid limit. Indeed, one of the key contributions of the present study is the theoretical elucidation of the non-trivial role of viscosity in the primary instability of tilted shear flows (see \S\ref{sec:scaling}), providing a more complete theoretical description on the recent experimental observation by \cite{Meunier2012}.} 

The paper is organised as follow. In \S \ref{sec:formulation}, the equations of motion are introduced and the linear stability analysis is formulated with its numerical method. In \S \ref{sec:scaling}, we simplify the set of linearised equations into an Orr-Sommerfeld type of equation in the limit of low Froude number. Based on the equation derived, we will provide a theoretical justification as to why the low-Froude-number instability would remain two dimensional while taking into account the length scale argument of \citet{Billant2001} and \citet{Brethouwer2007}. In \S \ref{sec:Result}, an absolute and convective instability analysis will be performed and its result will be compared with that of \citet{Meunier2012}. A further discussion will then be followed to address the issue of how the low-Froude-number mode is destabilised with decreasing Froude number. Finally, we will discuss what would be the expected nature of the base flow to explain the discrepancy between the experimental observation of \citet{Meunier2012} and the stability analysis of \citet{Candelier2011}. A summary and concluding remarks of this paper will be given in \S\ref{sec:conclusion}.

\section{Problem formulation \label{sec:formulation}}
\subsection{Equations of motion}

\begin{figure}
    \centering
    \includegraphics[width=0.7\textwidth]{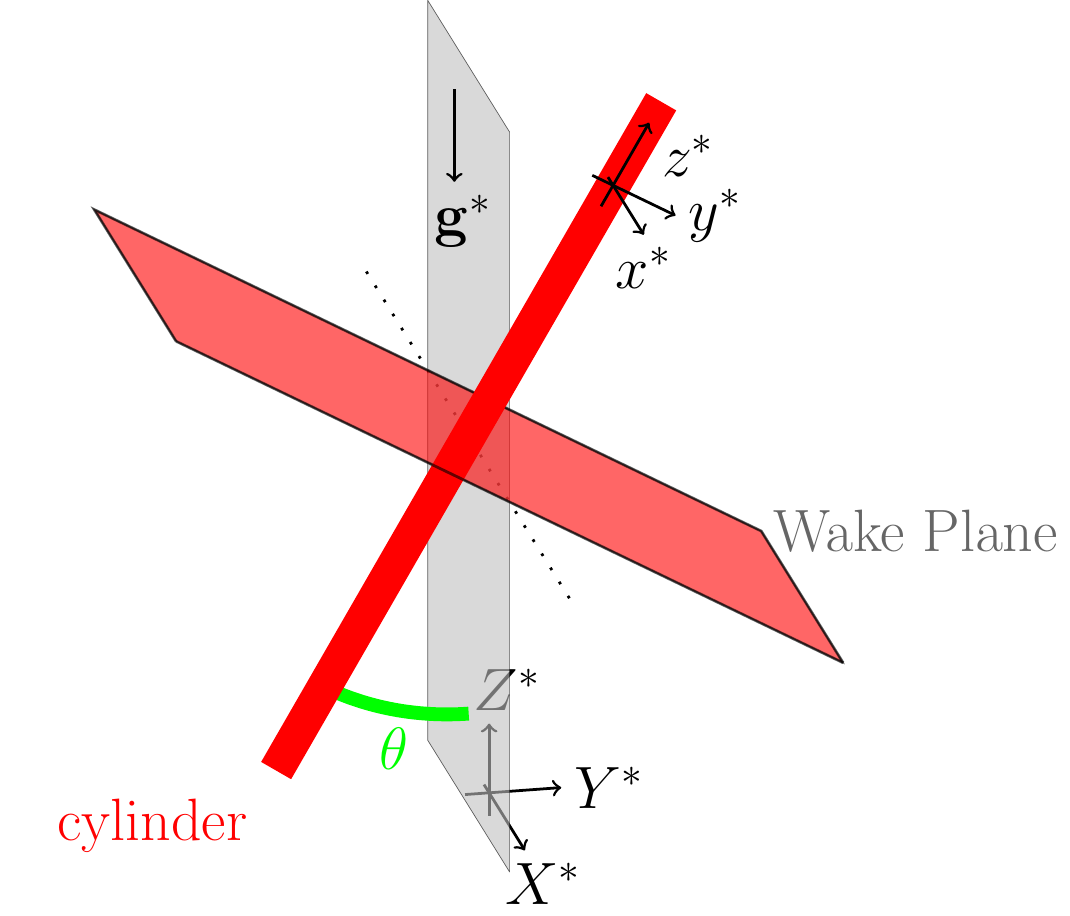}
    \caption{Sketch of the flow configuration and the coordinate systems. \label{fig:fig1}}
\end{figure}

Given the flow configuration where the base flow is tilted against the direction of gravity, it is instructive to start by introducing the coordinate systems used in the present study. We will adopt the same Cartesian coordinate systems as those in \citet{Meunier2012}, as is illustrated in figure \ref{fig:fig1}. Here, $(x^*,y^*,z^*)$ are the coordinates aligned with a two-dimensional bluff body (i.e. cylinder) with $x^*$ being the streamwise, $y^*$ the transverse, $z^*$ the spanwise direction, respectively. (Note that, throughout the present study, the superscript $^{*}$ indicates dimensional quantities, while those without it are non-dimensionalised ones.) The $(x^*,y^*,z^*)$ coordinate system is set to be tilted against the laboratory one defined by $(X^*,Y^*,Z^*)$ coordinates with an angle $\theta$. We shall assume that the base-flow (i.e. wake) profile remains to be unchanged in the $(x^*,y^*,z^*)$ coordinates, although this issue will be discussed later in \S \ref{subsec:corrections}. The relation between the two coordinate systems is then written as 
\begin{equation}
    (X^*,Y^*,Z^*)=(x^*,y^*\cos\theta+z^*\sin\theta,-y^*\sin\theta+z^*\cos\theta).
\end{equation}
In the $(x^*,y^*,z^*)$ coordinate, velocity is denoted by $\boldsymbol{u}^*=(u^*,v^*,w^*)$ and pressure by $p^*$. Similarly, in the $(X^*,Y^*,Z^*)$ coordinates, velocity is denoted by $\boldsymbol{U}^*=(U^*,V^*,W^*)$. Finally, the gravity is acting in the negative $Z^*$ direction, such that the density variation is imposed along the same direction. 

Under the Boussinesq approximation, the dimensionless equations of motion, defined in the $(x,y,z)$ coordinate system, are given as follows:
\begin{subequations}
    \begin{equation}
        \frac{\partial\boldsymbol{u}}{\partial t}+({\boldsymbol{u}}\bcdot \bnabla) \boldsymbol{u}=-\bnabla p+\frac{1}{\Rey} \nabla^2 \boldsymbol{u}+b\hat{\boldsymbol{g}},\label{eq:non_dim_u}
    \end{equation}
    \begin{equation}
        \frac{\partial b}{\partial t}+(\boldsymbol{u}\bcdot\boldsymbol{\nabla})b=\frac{1}{\Rey \Sc} \nabla^2 b,\label{eq:non_dim_rho}
    \end{equation}
\end{subequations}
in which $\Rey={U}_{ref}^{*}D^{*}/\nu^{*}$ is the Reynolds number, $\Sc=\nu^{*}/\kappa^{*}$ the Schmidt number, $p$ the pressure, and $\hat{\boldsymbol{g}}$ the unit vector representing the direction of gravity. Here, $\nu^{*}$ is the kinematic viscosity, $\kappa^{*}$ the diffusivity, and $g^{*}$ the gravitational acceleration, while the reference length scale $D^*$ and the velocity scale ${U}_{ref}^*$ will be defined later with the introduction of the base-flow profile. For the density fluctuation, we consider the dimensionless buoyancy $b$, defined as $(g^{*}D^{*}/{U}_{ref}^{*2})(\rho/\rho_0)$, where $\rho$ is the non-dimensional density and $\rho_0=\rho|_{y=0}$ is the dimensionless density at the origin. Here, $\rho|_{y=0}$ is set to equal to unity because $\rho^*|_{y=0}$ is used as the reference density for non-dimensionalisation.

A two-dimensional parallel wake is considered as an example of tilted shear flow. The wake flow is chosen to make a comparison with the experimental result in \citet{Meunier2012}, although we shall see that most of the arguments made in this paper are not restricted to the wake flow. The base-flow profile of the parallel wake in the present study is identical to the one in \citet{Monkewitz1988}:
\begin{subequations}\label{eq:U_profile}
    \begin{equation}
        u_0(y;R,a)=1-R+2RF(y),
    \end{equation}
with
    \begin{equation}
        F(y)=\{1+\sinh^{2a}(y\,\mathrm{arcsinh}1)\}^{-1},
    \end{equation}
\end{subequations}
in which $u_0(y)$ is the streamwise velocity of the base flow, $a$ the shape parameter (or the stiffness parameter), $R$ the
velocity ratio, defined by $R=(u_{c}-u_{\infty})/(u_{c}+u_{\infty})$, where $u_{c}$ and $u_{\infty}$ are the centreline and the freestream velocities, respectively. If $a\rightarrow\infty$, the profile becomes a top hat, whereas if $a$ is small, the profile becomes very smooth. Also, for  $-1<R<0$, the profile depicts a wake with no counter flow, while, for $R<-1$, it becomes a wake with counter flow at the centre. From the base-flow profile in (\ref{eq:U_profile}), the reference velocity is defined as ${U}_{ref}^{*}=(u_{c}^{*}+u_{\infty}^{*})/2$. The reference length scale $D^*$ is defined to be $u_0^{*}(y^*=D^{*})={U}_{ref}^{*}$ ($u_0(y=1)=1$ equivalently). Finally, for the buoyancy, $b$, a stable linear stratification is considered as the basic state: i.e. $b_0=(g^{*}D^{*}/{U}_{ref}^{*2})(1+(d\rho/dZ)Z)$ where $d\rho/dZ$ is constant.

Now, we consider a small perturbation around the basic state:
\begin{equation}
    \qquad\boldsymbol{u}=\boldsymbol{u}_0(y)+\boldsymbol{u}',\qquad b=b_0+b',\label{eq:Perturb}
\end{equation}
where $'$ represents the perturbation variables, and $\boldsymbol{u_0}(y)=(u_0(y),0,0)$. Then, the linearised equations of motion are given as 
\begin{subequations}\label{eq:lineareq}
    \begin{equation}
        \frac{\partial\boldsymbol{u}'}{\partial t} +(\boldsymbol{u}_0\bcdot\bnabla)\boldsymbol{u}' +(\boldsymbol{u}'\bcdot\bnabla)\boldsymbol{u}_0 =-\bnabla p'+ \frac{1}{\Rey} \nabla^2 \boldsymbol{u}' +b'\hat{\boldsymbol{g}},\label{eq:non_dim_u_Bos}
    \end{equation}
    \begin{equation}
        \frac{\partial b'}{\partial t} +(\boldsymbol{u}_0 \bcdot\bnabla)b' -(\boldsymbol{u}'\bcdot\bnabla)(Z/\Frou^{2}) =\frac{1}{\Rey \Sc}\nabla^2 b',\label{eq:non_dim_b}
    \end{equation}
\end{subequations}
where $\Frou=U_{ref}^*/(N^*D^*)$ with $N^*$ being the constant Brunt-V\"{a}is\"{a}l\"{a} frequency. Equations (\ref{eq:lineareq}) then admit the following normal-mode solution:
\begin{equation}
    \left[\begin{array}{c}
        u'\\
        v'\\
        w'\\
        b'\\
        p'
    \end{array}\right]=\left[\begin{array}{c}
        \tilde{u}(y)\\
        \tilde{v}(y)\\
        \tilde{w}(y)\\
        \tilde{b}(y)\\
        \tilde{p}(y)
    \end{array}\right]\exp\{i(\alpha x+\beta z-\omega t)\},\label{eq:pert_Fourier}
\end{equation}
where $\alpha$ and $\beta$ are given real wavenumbers in the $x$ and $z$ directions, and $\omega$ is the complex frequency. We can then write (\ref{eq:lineareq}) and the continuity equation as:
\begin{subequations}\label{eq:normalmode}
    \begin{equation}
        i\omega \tilde{u}=\mathcal{L}\tilde{u}+DU\tilde{v}+i\alpha\tilde{p},\label{eq:lin_u}
    \end{equation}
    \begin{equation}
        i\omega \tilde{v}=\mathcal{L}\tilde{v}+D\tilde{p}-\tilde{b}\sin\theta,\label{eq:lin_v}
    \end{equation}
    \begin{equation}
        i\omega \tilde{w}=\mathcal{L}\tilde{w}+i\beta\tilde{p}+\tilde{b}\cos\theta,\label{eq:lin_w}
    \end{equation}
    \begin{equation}
        i\omega \tilde{b}=\mathcal{L}_{\rho}\tilde{b}+\frac{\sin\theta}{\Frou^{2}}\tilde{v}-\frac{\cos\theta}{\Frou^{2}}\tilde{w},\label{eq:lin_rho}
    \end{equation}
    \begin{equation}
        i\alpha\tilde{u}+D\tilde{v}+i\beta\tilde{w}=0,\label{eq:lin_incompress}
    \end{equation}
\end{subequations}
where $D={d}/{dy}$, $k^{2}=\alpha^{2}+\beta^{2}$, $\mathcal{L}=iu_0\alpha-(D^2-k^2)/\Rey$ and $\mathcal{L}_{\rho}=iu_0\alpha-(D^2-k^2)/\Rey \Sc$.

\subsection{Numerical Method}

The equations (\ref{eq:normalmode}) are solved as an eigenvalue problem, in which $\omega$ becomes the eigenvalue and $(\tilde{u},\tilde{v},\tilde{w},\tilde{b},\tilde{p})^{T}$
is the corresponding eigenfunction. We discretise (\ref{eq:normalmode})
using a Chebyshev collocation method \citep{Weideman2000}, and solve
the resulting numerical eigenvalue problem with the \texttt{eig} function in
the \textsc{MATLAB} library. All of the
following results are computed \replaced{up to 300}{with 100} mesh points with the wall-normal domain size
of $y\in[-60.6,60.6]$\added{ -- such a large number of grid points was needed for three-dimensional instability mode emerging at low $Fr$ with non-zero tilting angle.} 
\added{Zero velocity perturbations and zero buoyancy fluctuation flux are imposed at the boundaries.} \deleted{We have checked the result using 300 mesh points, and it does not yield any discernible difference.} The numerical solver is also validated by comparing with the results in \citet{Hazel1972} by setting $\theta=0$ and with those in \citet{Candelier2011} by setting $\Rey\rightarrow \infty$.


In the present study, mainly $\Rey$, $\Frou$, tilting angle $\theta$ are varied, while keeping $Sc$ fixed at a value of 700. However, keeping $Sc$ as the same value is not a great limitation\deleted{, as was previously discussed by \mbox{\citet{Meunier2012}}}. \added{Indeed, the change of $Sc$ was found not to yield any discernible behaviour of the flow as long as it is large enough.} The present analysis is performed for the wake velocity profile at $a=1.34$ and $R=-1.105$ \citep{Monkewitz1988} defined in the $(x,y,z)$ coordinates, and the effect of the wake velocity profile is discussed in \S \ref{subsec:AU} and \S \ref{subsec:corrections}.

\section{Scaling analysis \label{sec:scaling}}

\subsection{A low Froude number approximation \label{subsec:Asymptotic-argument}}

Before proceeding to the numerical result of the stability analysis, we first examine the equations of motion in the low Froude number limit to explore any possible instability process. 

\begin{figure}
    \centering
    \includegraphics[width=0.9\textwidth]{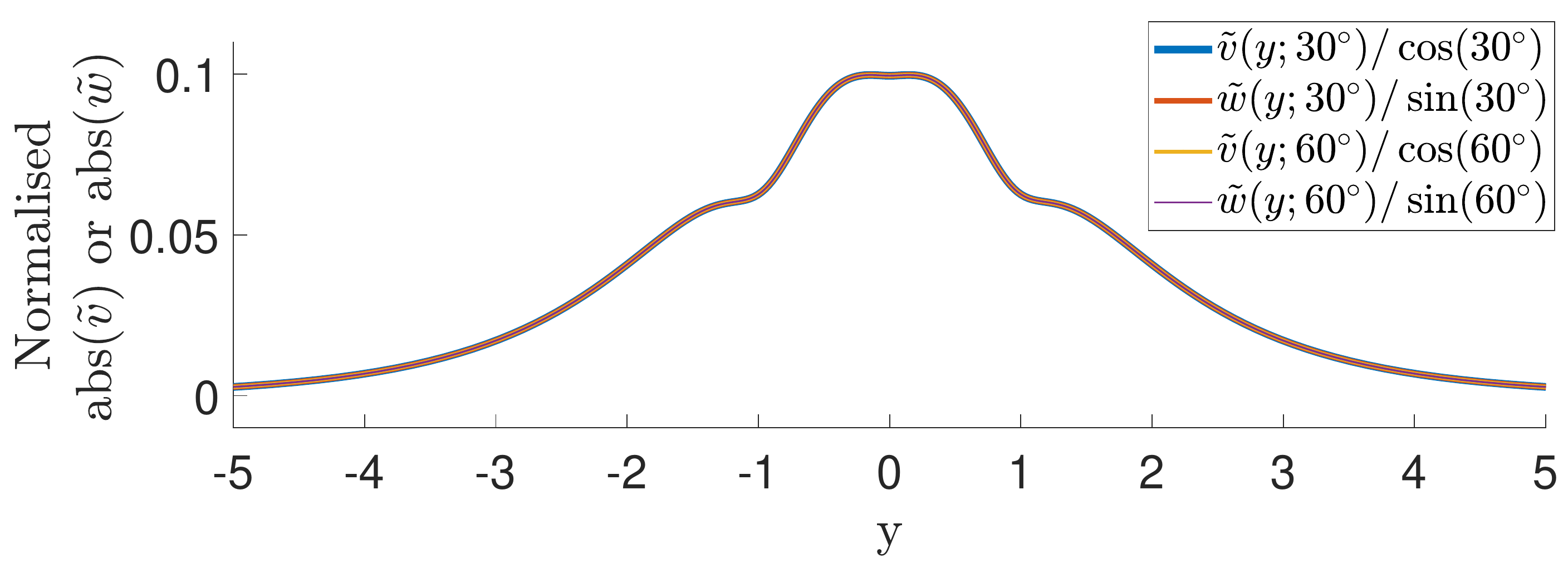}
    \caption{Confirmation of the relation (\ref{eq:v_w_same}) at $\theta=30^{\circ}$ and $60^{\circ}$ using numerically calculated eigenmodes ($\Frou=0.01$ and $\Rey\rightarrow \infty$). \label{fig:fig2}}
\end{figure}

The equation (\ref{eq:lin_rho}), containing $\Frou$, can be written as
\begin{equation}
    v'\sin\theta=w'\cos\theta -\Frou^{2}[\frac{\partial}{\partial t}+u_0 \frac{\partial}{\partial x}-\frac{\nabla^2}{\Rey \Sc}]b'. \label{eq:v_w_same_dev}
\end{equation}
If we take $\Frou \rightarrow 0$ and $\Frou^2/\Rey \Sc \rightarrow 0$ with the assumption of finite $\Rey$, equation (\ref{eq:v_w_same_dev}) yields
\begin{equation}
    w'\cos\theta-v'\sin\theta=0. \label{eq:v_w_same}
\end{equation}
Now, it is not difficult to realise that the \replaced{left}{right}-hand side of (\ref{eq:v_w_same}) is simply the vertical velocity fluctuation $W'$ in the $(X,Y,Z)$ coordinates, indicating $W'=0$. 
Numerical results at $\Frou=0.01$ also confirms this observation, as shown in figure \ref{fig:fig2}. The result of (\ref{eq:v_w_same}) indicates the suppression of vertical velocity by strong buoyancy force, such that the perturbation velocity field lies only on the horizontal plane. This suppression of the vertical velocity has also been well discussed in previous studies \citep[see][]{Candelier2011,Meunier2012}. Also, (\ref{eq:v_w_same_dev}) implies that $W'$ scales as $\Frou^2$ for small non-zero $\Frou$. For now, we shall proceed our discussion while keeping $W'=0$ in our approximation, and its validity will be discussed in \S \ref{subsec:vert_length_scale} in relation to the previously proposed asymptotic scaling \citep{Billant2001,Brethouwer2007}.

With (\ref{eq:v_w_same}), the $y$ and $z$ components of (\ref{eq:non_dim_u_Bos}) can be simplified into
\begin{equation}\label{eq:non_dim_pb}
    \sin\theta \frac{\partial p'}{\partial  y}- \cos\theta \frac{\partial p'}{\partial z}=b',
\end{equation}
implying that $w'$ and $b'$ in (\ref{eq:non_dim_u_Bos}) are explicitly given in terms of $v'$ and $p'$ with (\ref{eq:v_w_same}) and (\ref{eq:non_dim_pb}). Then, the momentum equation of (\ref{eq:lineareq}) can be written as the following single equation for $v'$:
\begin{subequations}\label{eq:v_OStot}
    \begin{equation}
        \left[(\frac{\partial}{\partial t}+u_0\frac{\partial}{\partial x}-\frac{\nabla^2}{\Rey})\nabla^2_\theta-\cos^2 \theta \frac{\partial^{2}u_0}{\partial y^{2}}\frac{\partial}{\partial x}\right]v'=0,\label{eq:v_OS}
    \end{equation}
where
    \begin{equation}
        \nabla^2_\theta=\pardxx{} + \cos^2 \theta \pardyy{} + 2 \cos \theta \sin \theta \pardyz{} + \sin^2 \theta \pardzz{}.
    \end{equation}
\end{subequations}
Here, it can be realised that $\nabla^2_\theta=\partial^2 / \partial X^2 + \partial^2 / \partial Y^2$ in the $(X,Y,Z)$ coordinates, indicating that (\ref{eq:v_OS}) can be written as 
\begin{equation}
    \left[\left( \frac{\partial}{\partial t}+U_0\frac{\partial}{\partial X}-\frac{1}{\Rey}(\frac{\partial^{2}}{\partial X^{2}}+\frac{\partial^{2}}{\partial Y^{2}}+\frac{\partial^{2}}{\partial Z^{2}}) \right) (\frac{\partial^{2}}{\partial X^{2}}+\frac{\partial^{2}}{\partial Y^{2}})-\frac{\partial^{2}U_0}{\partial Y^{2}}\frac{\partial}{\partial X}\right]V'=0.\label{eq:v_lowFr_No_Modal}
\end{equation}

From (\ref{eq:v_lowFr_No_Modal}), we can make several important observations on the nature of instabilities arising in (\ref{eq:lineareq}) when $\Frou \rightarrow 0$:
\begin{enumerate}
    \item Equation (\ref{eq:v_lowFr_No_Modal}) is no more dependent of $\Frou$ and $b'$, implying that the density stratification cannot affect any instability arising from (\ref{eq:v_lowFr_No_Modal}). Furthermore, (\ref{eq:v_lowFr_No_Modal}) only contains the horizontal shear $\partial U_0/\partial Y$, suggesting the barotropic nature of the possible instability at $\Frou \rightarrow 0$ (i.e. instability in the horizontal plane). 
    \item The form of (\ref{eq:v_lowFr_No_Modal}) is very similar to the physical-space Orr-Sommerfeld equation in the $X$-$Y$ plane: in fact, it is identical to the Orr-Sommerfeld equation, except the term with $\partial^{2}/\partial Z^{2}$ in (\ref{eq:v_lowFr_No_Modal}). Such a similarity strongly suggests that the given horizontal shear would admit an inflectional instability if $\partial^{2}U_0 / \partial Y^{2}$ is not zero for some $Y$. 
    \item In (\ref{eq:v_lowFr_No_Modal}), $\partial^{2}/\partial Z^{2}$ emerges in the viscous term, implying that any vertical variation in the velocity perturbation might be stabilising via viscous dissipation. Indeed, we shall see the purely stabilising effect of the $\partial^{2}/\partial Z^{2}$ term in \S \ref{subsec:Squire}. This observation also has important implication to the vertical length scale, as we shall see in \S \ref{subsec:vert_length_scale}. 
    \item  Finally, in (\ref{eq:v_lowFr_No_Modal}), there is no explicit dependence on the tilting angle $\theta$ because all the $\theta$-dependent terms disappear by introducing $(X,Y,Z)$ coordinates. This indicates that the stability of (\ref{eq:lineareq}) at $\Frou \rightarrow 0$ would mainly be affected by the horizontal projection of the given base-flow shear (i.e. $\partial^{2}U_0 / \partial Y^{2}$ in (\ref{eq:v_lowFr_No_Modal})). From this observation, we shall discuss the implication on how the tiliting angle $\theta$ affects the stability at $\Frou \rightarrow 0$ in \S \ref{subsec:corrections}. 
\end{enumerate}
\subsection{\added{Primary instability of tiltiled shear flows at low $\Frou$} \label{subsec:2D}}
\added{As suggested by \citet{Deloncle2007} and \citet{Candelier2011}, there is no Squire's theorem stating that the most unstable mode is two-dimensional in horizontal and tilted shear flows. However, the numerical results of \citet{Deloncle2007} for inviscid horizontal shear flows and \citet{Candelier2011} for inviscid tilted Bickley jets showed that the most unstable mode arises when $\beta=0$ (i.e. when the mode is two-dimensional). The same is numerically true in the present viscous case, as we shall demonstrate both asymptotically and numerically.}
\subsubsection{Squire's theorem for weakly tilted shear flows \label{subsec:Squire}}
\deleted{As suggested by Deloncle et. al. (2007) and Candelier et al. (2011), there is no Squire's theorem stating that the most unstable mode is two-dimensional in horizontal and tilted shear flows. However, the numerical results of Deloncle et. al. (2007) for inviscid horizontal shear flows and Candelier et al. (2011) for inviscid tilted Bickley jets showed that the most unstable mode arises when $\beta=0$ (i.e. when the mode is two-dimensional). The same is numerically true in the present viscous case, as demonstrated in figure 3.}
In this section, instead of relying on such numerical calculations, we attempt to mathematically demonstrate that the two-dimensional instability mode is always the most unstable one for small $\theta$ as long as the flow belongs to the regime where the low Froude number approximation in \S\ref{subsec:Asymptotic-argument} is valid.  

The main technical difficulty here is that the spanwise direction in the $(x,y,z)$ coordinates is not orthogonal to the direction of gravity. However, the low Froude number approximation of the equation of motion (\ref{eq:v_lowFr_No_Modal}) relieves this difficulty. Let us consider a spanwise uniform mode (i.e. $\beta=0$) in the $(x,y,z)$ coordinates\deleted{, as sketched in figure (deleted)}. This mode may then be interpreted as a tilted stack of horizontal modes in the $(X,Y,Z)$ coordinates\deleted{, as shown in figure (deleted)}. If the given mode is uniform in the $z$-direction, each of these horizontal modes should have exactly the same spatial shape, but with a $Y$-direction shift depending on the vertical location $Z$. In other words, the perturbation velocity homogeneous in the $z$-direction should satisfy the following relation:
\begin{subequations}\label{eq:sym}
    \begin{equation}\label{eq:sym1}
        V'(X,Y,Z)=V'(X,Y-Z\tan\theta,0).
    \end{equation}
The same is true for the base flow $U_0(Y,Z)(=u_0(y))$ as it is uniform in the $z$-direction: i.e.
    \begin{equation}\label{eq:sym2}
        U_0(Y,Z)=U_0(Y-Z\tan\theta,0).
    \end{equation}
\end{subequations}


Now, we assume that the base flow is weakly tilted (i.e. $\theta \ll 1$), so that we can introduce $\epsilon_\theta=\tan \theta$ and $Z_0=\epsilon_\theta Z$. Using WKBJ approximation, we can write $V'(X,Y,Z)$ as 
\begin{equation}
    V'(X,Y,Z)=\tilde{V}(Y;Z_0) \exp \left[ \frac{i}{\epsilon_\theta} \int^{Z_0} k_Z(Z_0) d Z_0 -i\omega t + i k_X X \right],
\end{equation}
where $\omega$ is the eigenfrequency of the linear instability mode arising in the $(Y,Z)$ domain, $k_Z(Z_0)$ is the local wavenumber in the $Z$-direction and $k_X$ the streamwise wavenumber. At the leading order, (\ref{eq:v_lowFr_No_Modal}) becomes
\begin{equation}
    \left[(-i\omega+ik_X U_0-\frac{D_Y^2-k_X^2-k_Z^2(Z_0)}{\Rey})(D_Y^2-k_X^2)-ik_X D_Y^2 U_0\right]\tilde{V}(Y;Z_0)=0,\label{eq:V_OS_wavenumber_labframe}
\end{equation}
where $D_Y=\partial/\partial Y$. Here, we note that, if $k_z(Z_0)=0$, the stability property of (\ref{eq:V_OS_wavenumber_labframe}) does not change with $Z_0$ due to the nature of the base flow in (\ref{eq:sym2}). This then leads the eigenvalue $\omega$ in (\ref{eq:V_OS_wavenumber_labframe}) to satisfy the following relation:
\begin{equation}\label{eq:omega}
    \omega=\omega_{2D}-\frac{ik_z(Z_0)^2}{\Rey},
\end{equation}
where $\omega_{2D}$ is the eigenfrequency obtained from (\ref{eq:V_OS_wavenumber_labframe}) by setting $k_Z(Z_0)=0$. Now, it is evident that having non-zero $k_Z(Z_0)$ would only decrease the value of $\omega_i$. Therefore, it is not difficult to find that the most unstable mode globally arising in the $(Y,Z)$ domain is given when $k_Z(Z_0)=0$ for every $Z_0$. Since the linear operator in (\ref{eq:V_OS_wavenumber_labframe}) is invariant under the transformation of $(Y;Z_0) \rightarrow (Y-Z_0\tan\theta,0)$ for $k_Z(Z_0)=0$, the eigenmode obtained from (\ref{eq:V_OS_wavenumber_labframe}) would satisfy (\ref{eq:sym1}) for given $k_X$, indicating that the two-dimensional mode in the $(x,y,z)$ coordinate is the most unstable in a slightly tilted horizontal shear flow.
\deleted{We note that the argument here is valid only for small $\theta$ (i.e. slightly tilted case). For a strongly tilted case where the cylinder is close to horizontal, it may well break down, as the slowly varying assumption of the base flow in the $Z$-direction would not be valid any more. Indeed, Bosco \& Meunier (2014) has documented the appearance of another mode (Mode L) that is three-dimensional at high $\theta$. }

\subsubsection{\added{Numerical analysis for highly tilted shear flows}}
\floatsetup[figure]{style=plain,subcapbesideposition=top}
\begin{figure}
    \centering{}
	\sidesubfloat[]{
	    \includegraphics[width=0.45\columnwidth]{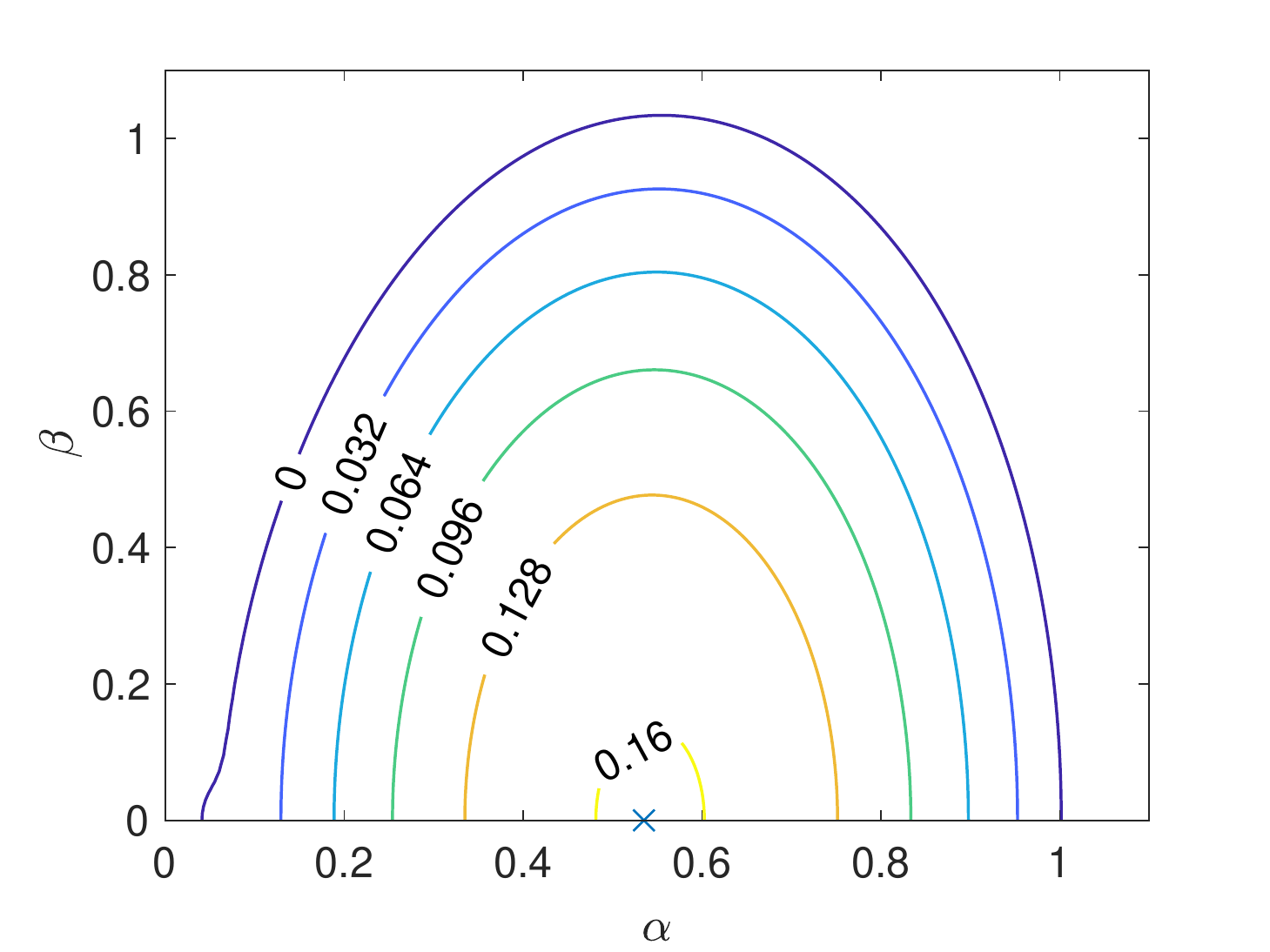}
    }
    \sidesubfloat[]{
        \includegraphics[width=0.45\columnwidth]{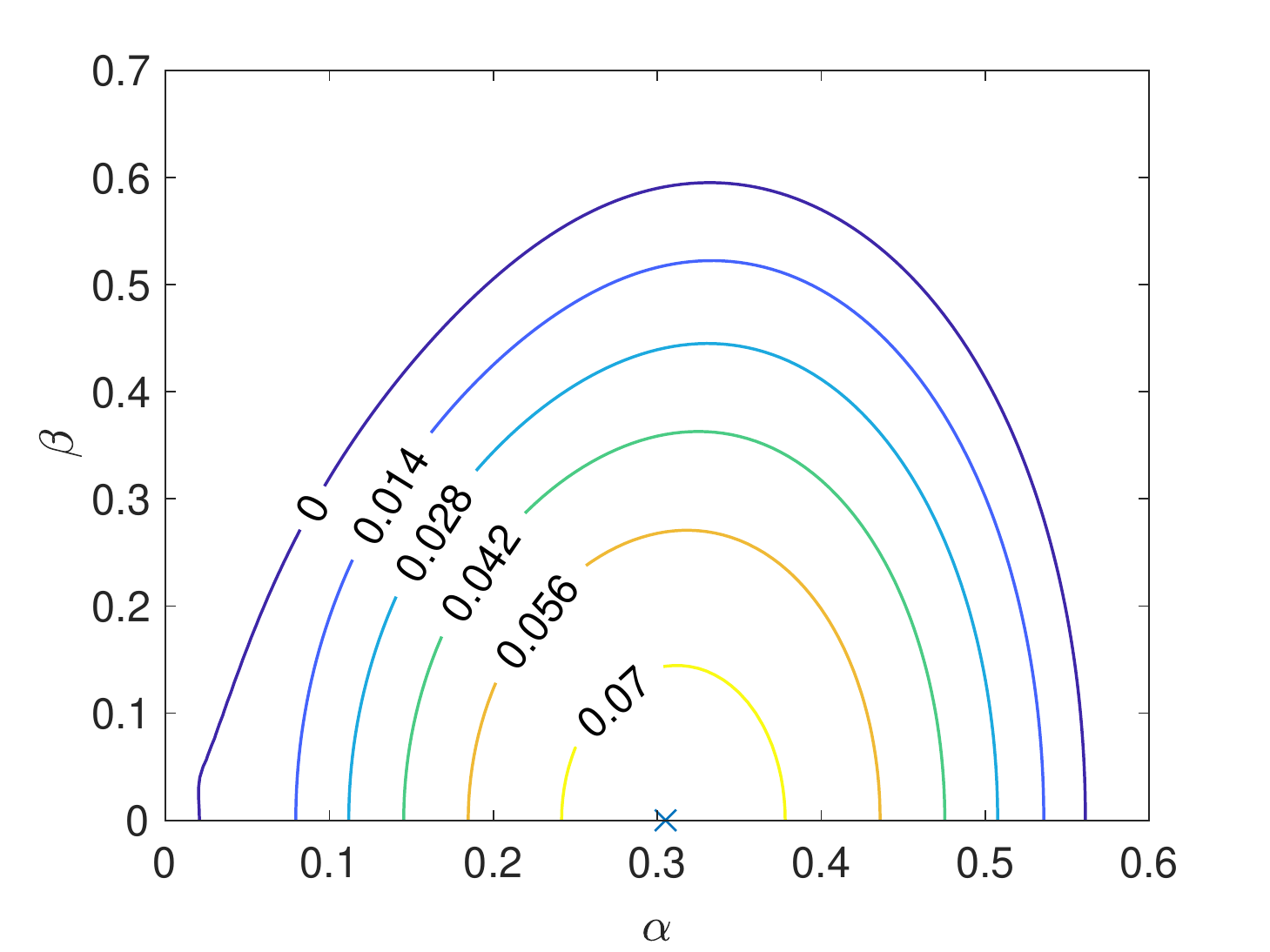}
    }\\
    \sidesubfloat[]{
        \includegraphics[width=0.45\columnwidth]{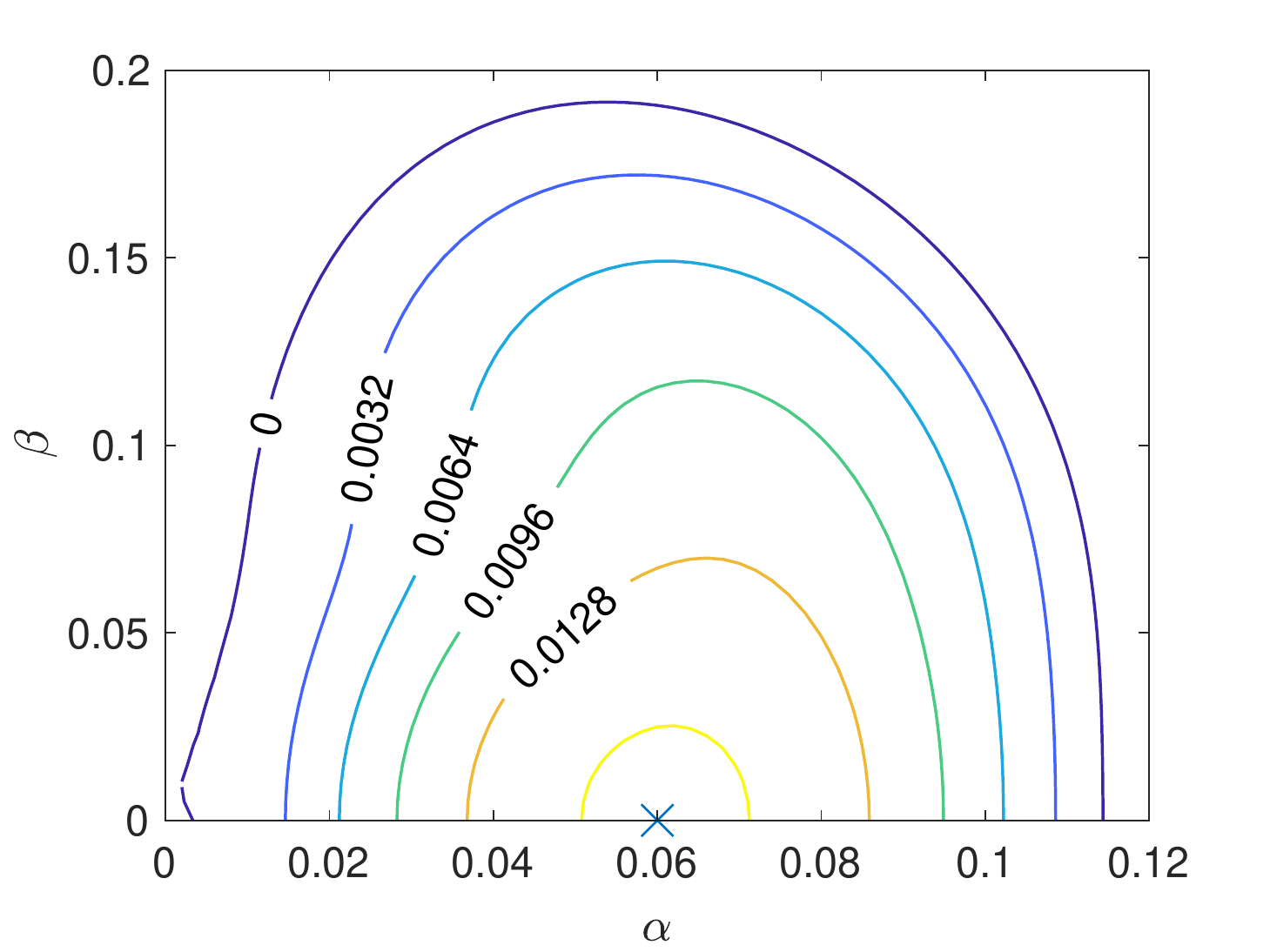}
    }
    \caption{Contour of the temporal growth rate ($\omega_{i,\max}$) of the most unstable mode in the real $\alpha-\beta$ plane \replaced{at $\Frou=0.01$, with $(a)$ $\Rey=7.8, \theta=30^\circ$, $(b)$ $\Rey=7.8, \theta=60^\circ$  $(c)$ $\Rey=50, \theta=85^\circ$. Here, the wavenumber of the most unstable mode is indicated with the cross symbol on each contour, and it is always given for $\beta=0$.}{($\Frou=0.01$, $\Rey=50$, and $\theta=60^{\circ}$). }\label{fig:fig4}}
\end{figure}
\added{We note that the theoretical argument in \S \ref{subsec:Squire} is valid only for small $\theta$ (i.e. slightly tilted case). For a strongly tilted case where the cylinder is close to horizontal, it does not guarantee the slowly varying assumption of the base flow in the $Z$-direction. However, the numerical result, as shown in figure \ref{fig:fig4}, reveals that the two-dimensional mode indeed remains to be most unstable even at tilting angle as high as $\theta=85^\circ$. Therefore, the numerical result extends the theoretical argument for weakly tilted flow made in the previous section to the strongly tilted one. This observation is also consistent with \cite{Meunier2012}, who experimentally showed that the horizontal vortex shedding emerges as the primary instability for any titling angles.}

\subsubsection{\added{Numerical analysis for higher Froude number}}
\floatsetup[figure]{style=plain,subcapbesideposition=top}
\begin{figure}
    \centering{}
	\sidesubfloat[]{
	    \includegraphics[width=0.45\columnwidth]{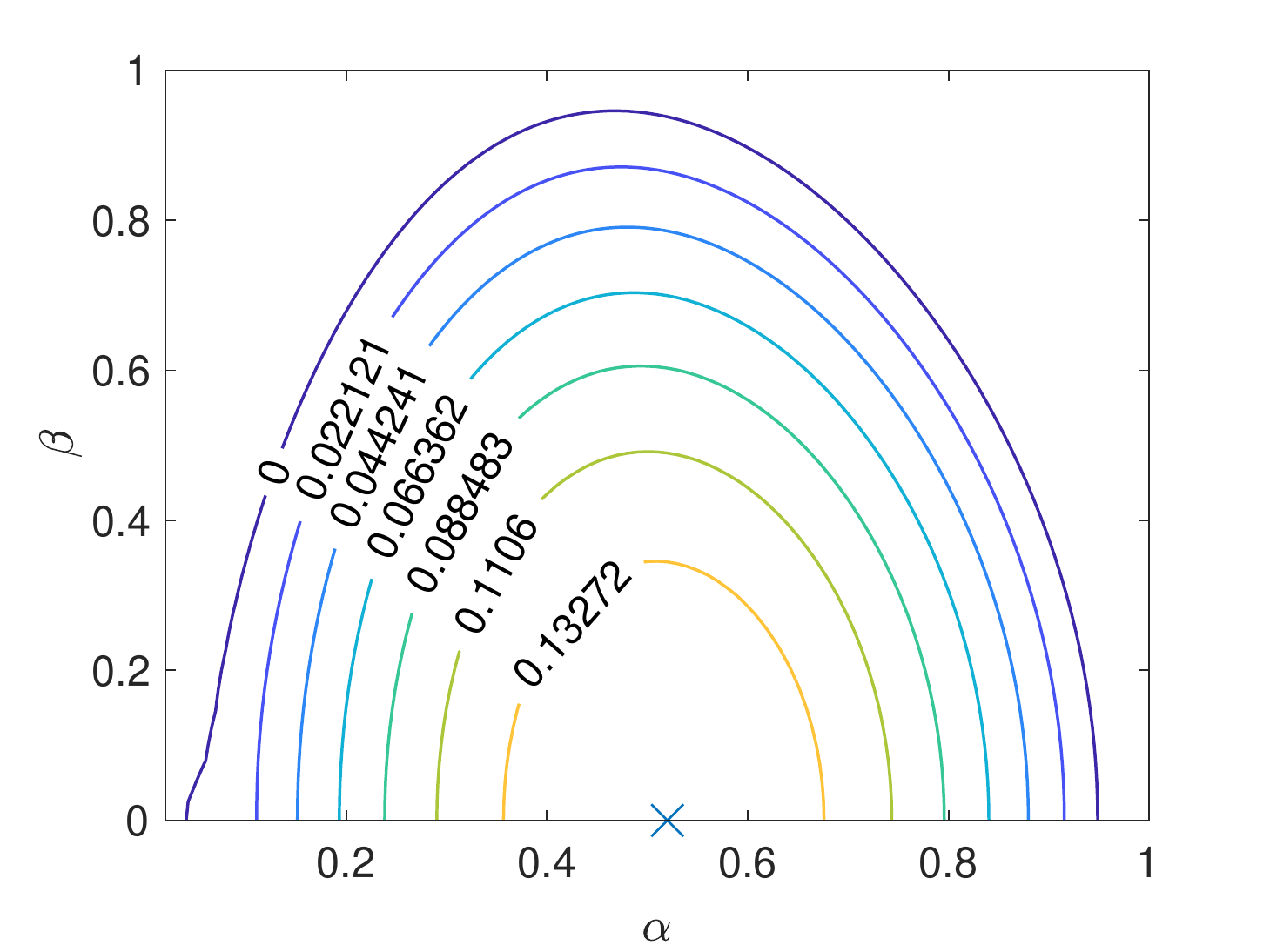} 
    }
    \sidesubfloat[]{
        \includegraphics[width=0.45\columnwidth]{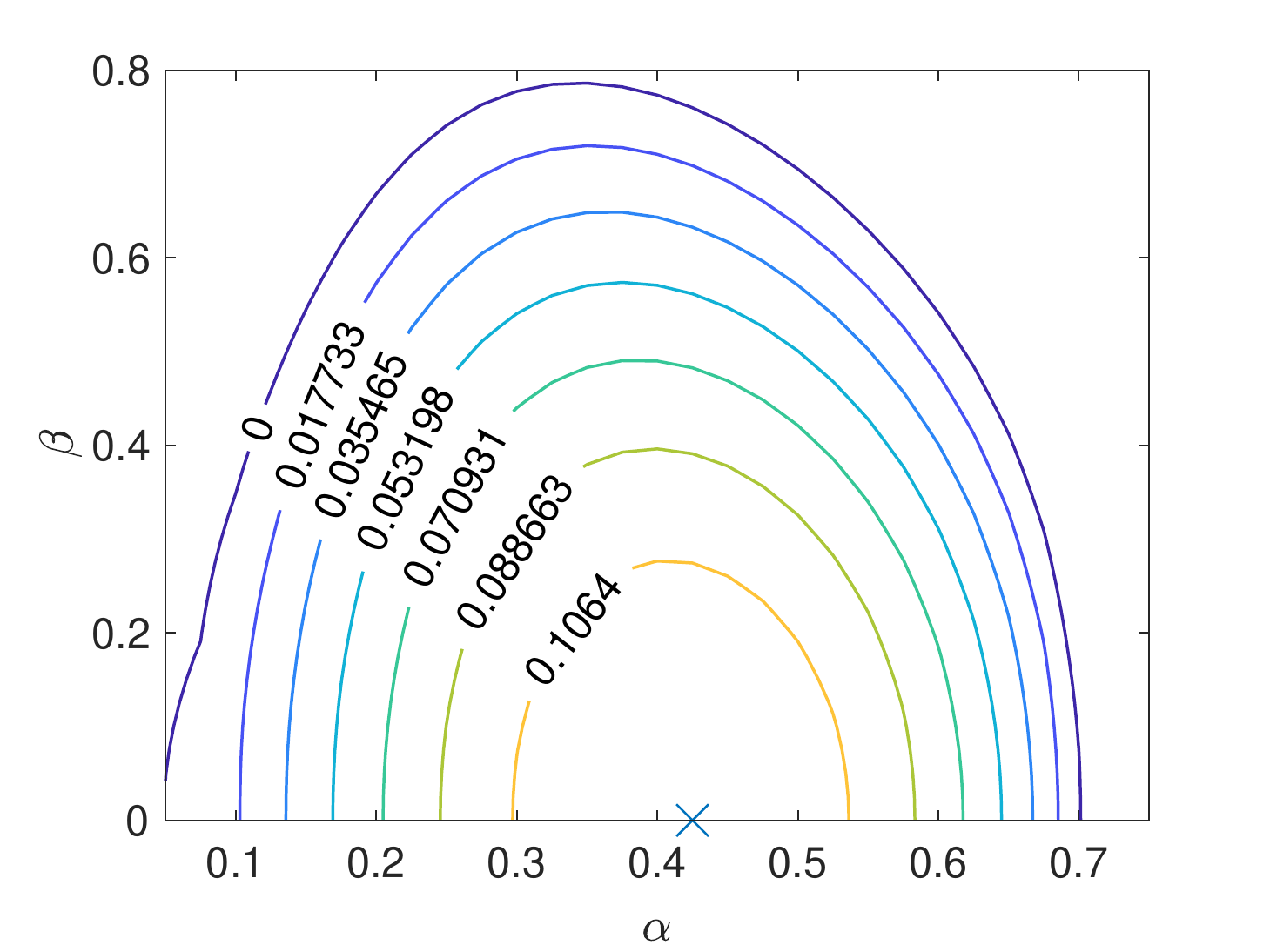} 
    }\\
	\sidesubfloat[]{
	    \includegraphics[width=0.45\columnwidth]{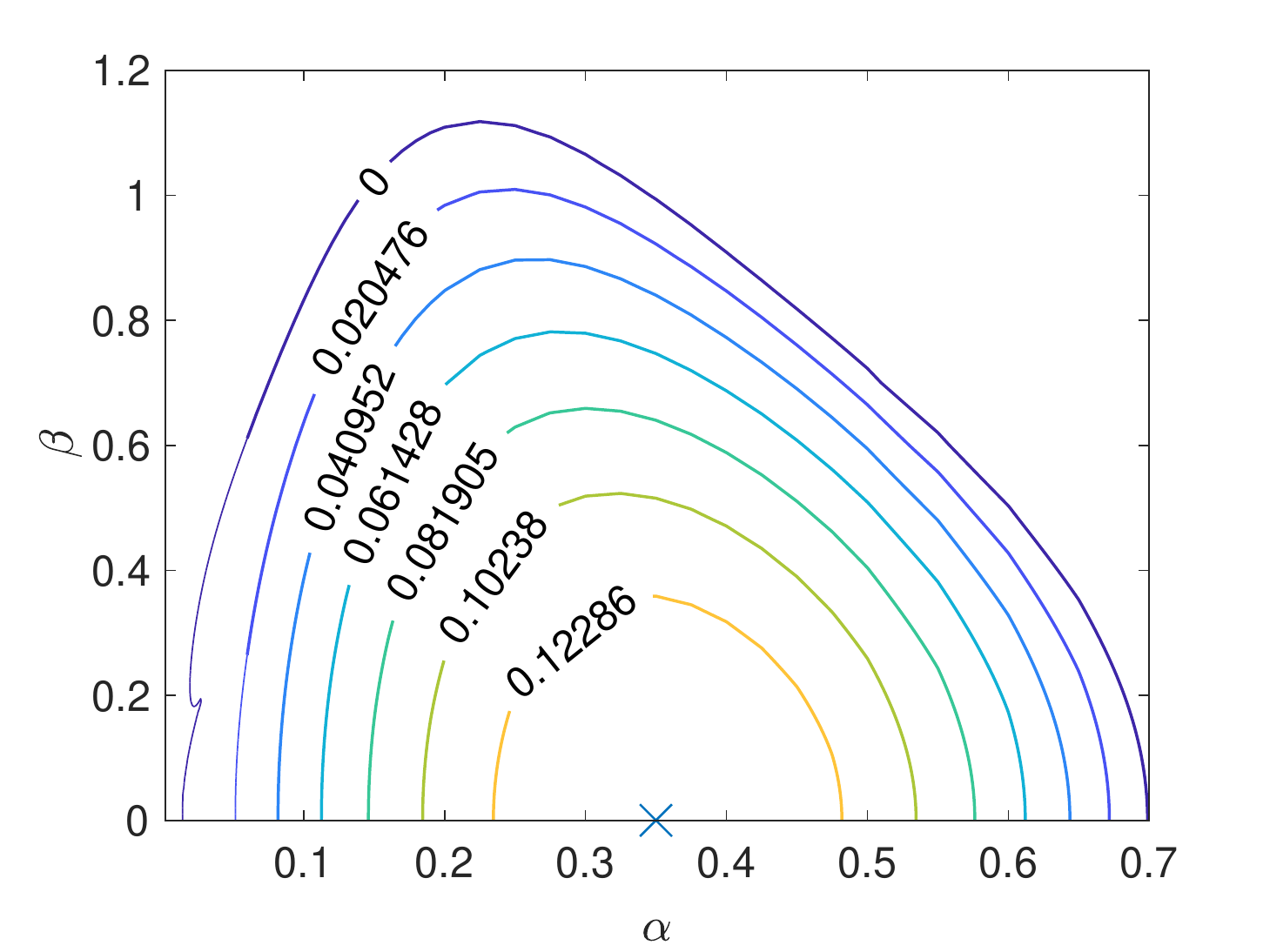}
    }
    \sidesubfloat[]{
        \includegraphics[width=0.45\columnwidth]{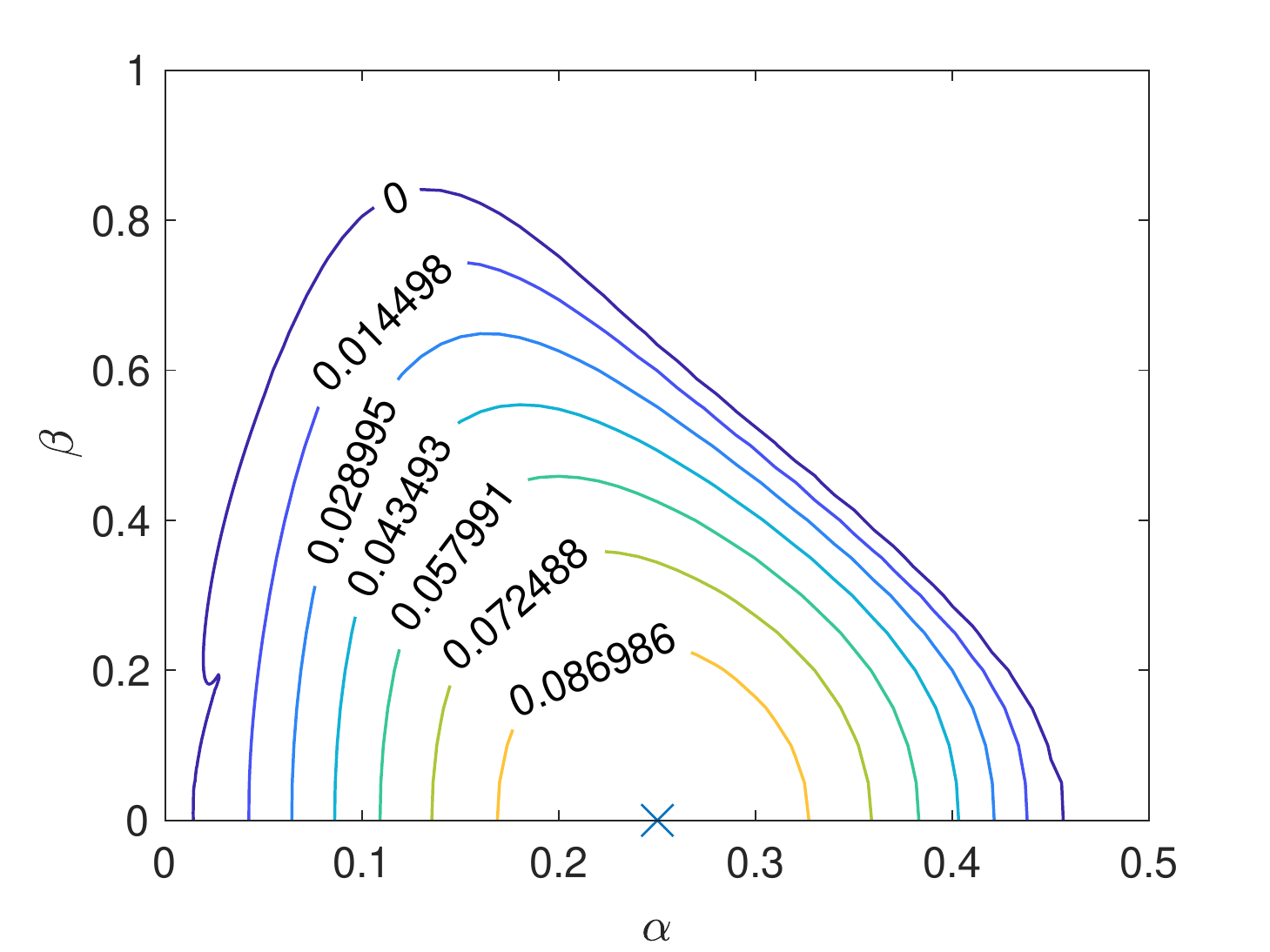} 
    }
    \caption{\added{Contour of the temporal growth rate ($\omega_{i,\max}$) of the most unstable mode in the real $\alpha-\beta$ plane at $(a,b)$ $\Rey=7.8$ and $(c,d)$ $\Rey=50$, with $(a,c)$ $\Frou=0.5$ and $(b,d)$ $\Frou=1$. Here, $\theta$ is $(a,b)$ $30^\circ$ and $(c,d)$ $60^\circ$. Here, the wavenumber of the most unstable mode is indicated with the cross symbol on each contour, and it is always given for $\beta=0$.}\label{fig:fig4ext}}
\end{figure}
\added{We have also extended the numerical result to higher $\Frou$ for completeness, as shown in figure \ref{fig:fig4ext}. Although there seems to be another emerging three-dimensional mode at low $\alpha$ when $\theta=60^\circ$ and $\Rey=50$ (see the kinks in figures \ref{fig:fig4ext}$c$, $d$, which indicate a branch-switching behaviour), the two-dimensional inflectional instability mode still remains to be most unstable, consistent with the observation of \citet{Meunier2012}. However, in this regime of parameters (i.e. relatively high buoyancy Reynolds number), it is important to note that the Squire-like theorem, which we demonstrated previously, does not precisely apply, as will be discussed in \S\ref{subsec:vert_length_scale}. Indeed, the recent work by \citet{Facchini2018a} has shown that a three-dimensional instability can arise in horizontal plane Couette flow where inflectional instability mechanism is ruled out by its base flow. While we have not observed such a three-dimensional instability as the most unstable primary instability in the present study, we do not rule out such possibility in other flow configurations where buoyancy Reynolds number is not small.
}

\subsection{\added{Asymptotic regimes, vertical length scales and primary instabilities}\label{subsec:vert_length_scale}}
The highly horizontal nature (i.e. very small vertical velocity) of the possible instability mode at $\Frou \rightarrow 0$ in the present study might remind one of the features of pancake vortical structures in a typical geophysical flow setting. However, given our discussion in \S\ref{subsec:Squire}, the primary instability in the present study would not vary vertically, unlike the pancake vortical structures. In fact, the key difference between the regime of the present analysis and the one of geophysical flow originates from the strength of viscosity. In the present study, the vertical derivative in (\ref{eq:v_lowFr_No_Modal}) appears in the viscous dissipation term, implying that the vertical direction would be correlated through viscous diffusion transport. Therefore, the appropriate vertical length scale in this case should be determined by viscosity. 
By contrast, in the geophysical flow regime where pancake vortical structures typically emerge, its vertical length scale would be determined by the strength of the stratification, as was proposed by \citet{Billant2001}.

The effect of viscosity in a strongly stratified medium has previously been discussed by \citet{Brethouwer2007}. They suggested the so-called buoyancy Reynolds number $\mathscr{R}= \Rey\Frou^2$ as the key parameter that determines the vertical length scale at low $\Frou$. If $\mathscr{R} \gg 1$, viscous force is unimportant and the relevant vertical length scale becomes proportional to $\Frou$ \citep{Billant2001}. On the other hand, if $\mathscr{R} \ll 1$ like in the present case, the vertical length scale would be proportional to $\Rey^{-1/2}$, the regime more relevant to typical laboratory experiments. 

\floatsetup[figure]{style=plain,subcapbesideposition=top}
\begin{figure}
	\centering{}
	\sidesubfloat[]{
	    \includegraphics[width=0.45\columnwidth]{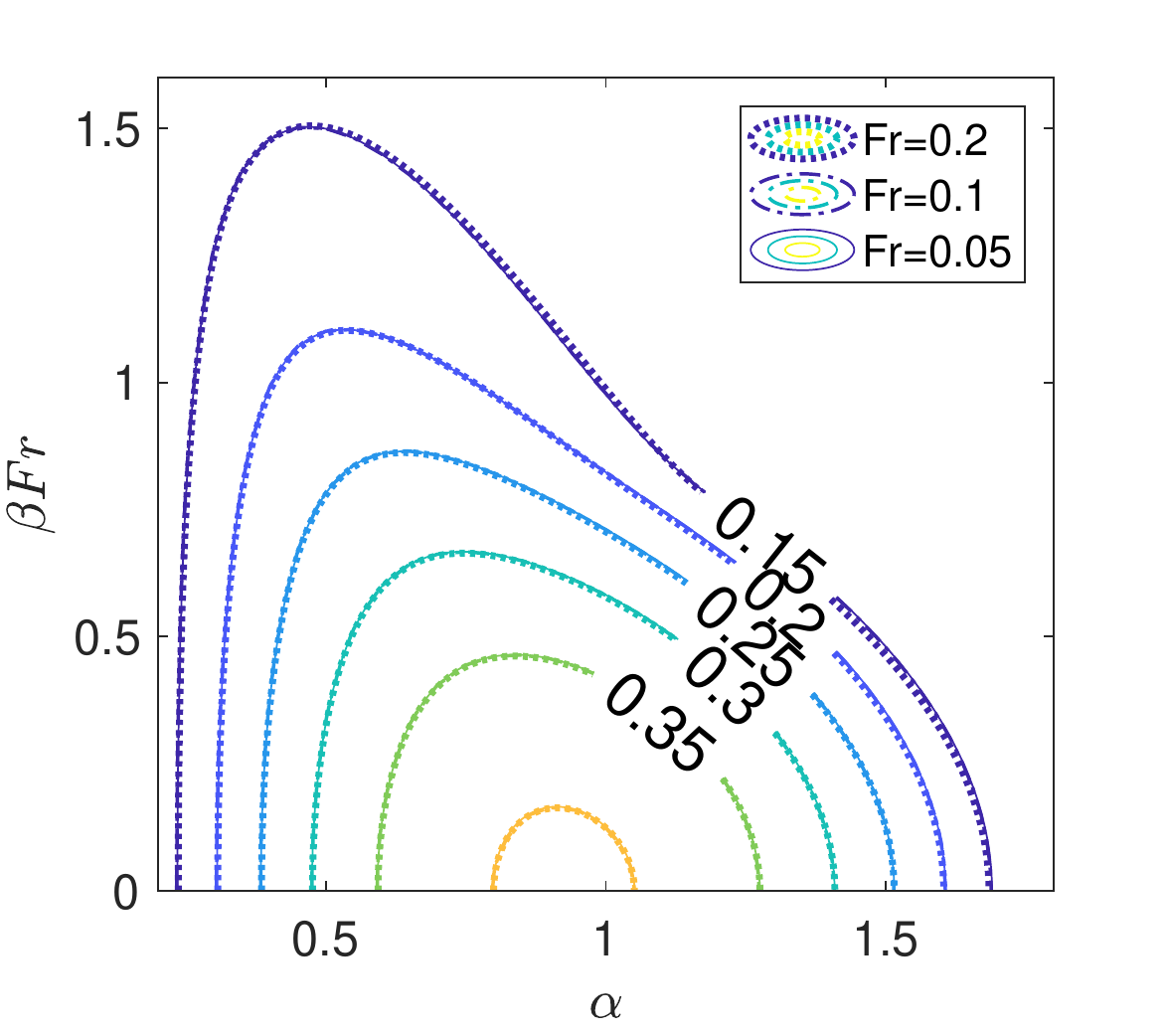}\label{fig:fig3a}
    }
    \sidesubfloat[]{
        \includegraphics[width=0.45\columnwidth]{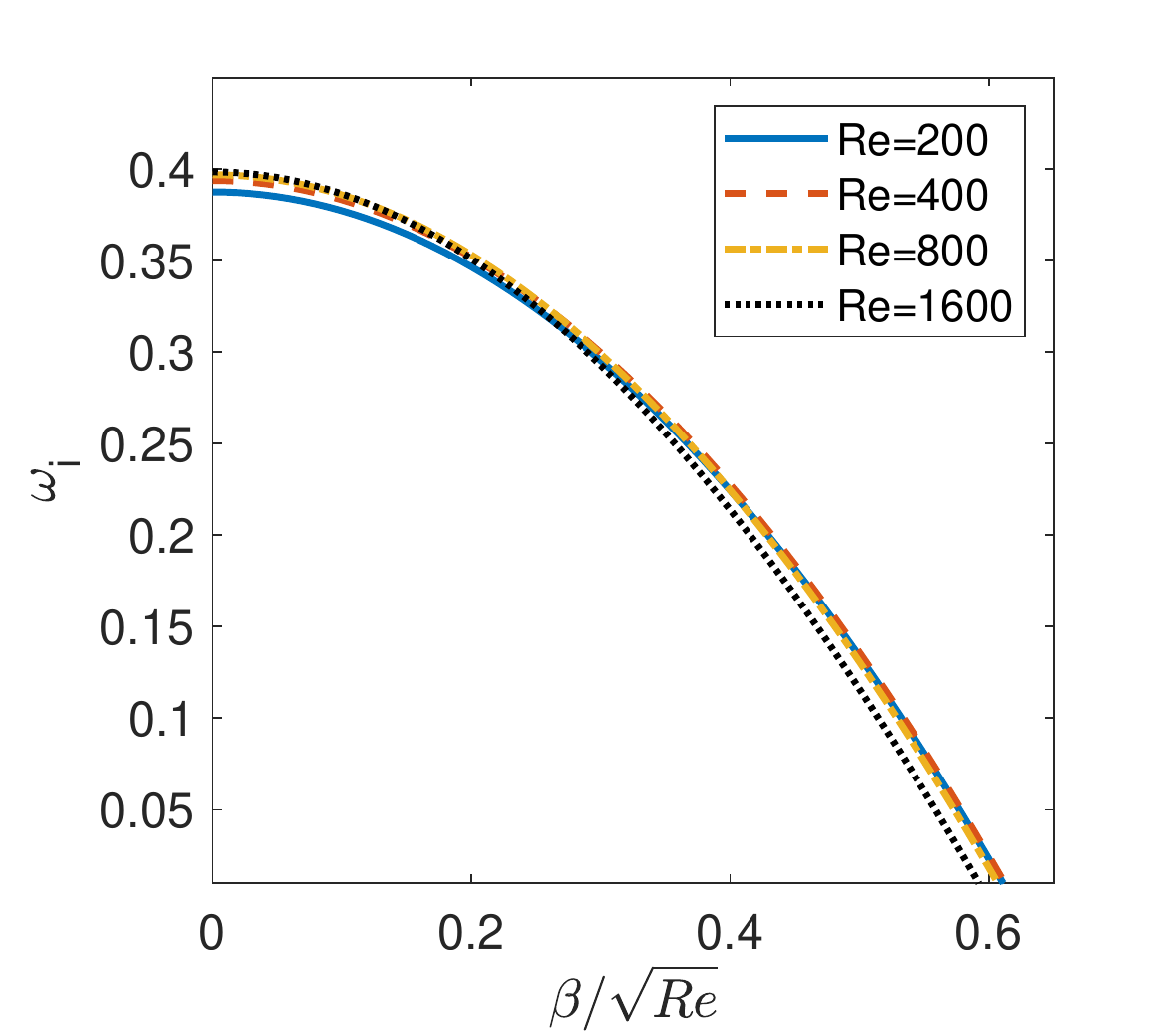}\label{fig:fig3b}
    }
    \caption{Self-similarity in the dispersion relation for purely horizontal flow ($\theta=0^\circ$): ($a$) contour of temporal growth rate $\omega_i$ in the $\alpha-\beta/\Frou$ plane for $\Frou=0.05,0.1,0.2$ ($\Rey \rightarrow \infty$); ($b$) temporal growth rate as a function of $\beta/\sqrt[]{\Rey}$ for $\Rey=200,400,800,1600$ ($\alpha=0.8$ and $\Frou=0.001$).\label{fig:fig3}}
\end{figure}

The argument on the relevant vertical length scale can also be demonstrated numerically in the present wake flow. Figure \ref{fig:fig3} presents a temporal stability analysis for a set of low Froude number in a wide range of Reynolds numbers when $\theta=0^\circ$. If the effect of viscosity is ignored to reach the regime of $\mathscr{R} \gg 1$, the growth rate shows self-similarity with respect to $\beta \Frou$ (figure \ref{fig:fig3a}) as was previously shown by \citet{Deloncle2007}. However, if the viscosity is brought into the analysis to be in the regime of $\mathscr{R} \ll 1$, the temporal growth rate exhibits self-similarity with respect to $\beta/\sqrt{\Rey}$ (figure \ref{fig:fig3b}), also consistent with (\ref{eq:omega}) (note that $\beta=k_Z(Z_0)$ for $\theta=0^\circ$). \added{In practice, $\mathscr{R} > O(10)$ is usually necessary to observe the layered pancake structures \citep{Lucas2017}}.

The numerical observation here and the asymptotic analysis in \S\ref{subsec:Squire} now suggest that if the primary instability in a strongly stratified horizontal shear flow emerges at sufficiently low Reynolds number, it is quite possible to be two dimensional. Here, it is important to remember that the typical inflectional instability arises at $\Rey\sim O(10)$ \citep{Ho1984}. Given the buoyancy Reynolds number argument, this implies that the primary instability, which originates from horizontal inflectional base flow, should be two dimensional as long as $\Frou$ \added{is below $O(1)$}. This also explains why the low-Froude-number instability mode observed in the experiment of \citet{Meunier2012} was two dimensional -- their $\Frou\sim O(10^{-1}\added{-1})$ and $\Rey \sim O(10)$, thus $\mathscr{R}$ \replaced{is below $O(10)$}{becomes well below unity}. In this respect, it is worth mentioning the recent work by \citet{Facchini2018a}\added{,} where a new type of three-dimensional linear instability was reported in horizontal Couette flow. However, in this case, $\Frou \sim O(10^{-1}-1)$ and $\Rey \added{> O(10^3)}$. Therefore, the buoyancy Reynolds number is $\mathscr{R} \sim O(\added{10^2-10^4})$, which does not belong to the asymptotic regime discussed in \replaced{this section}{ \S\ref{subsec:Asymptotic-argument} and \S3.2}. 

\deleted{Lastly, the above self-similarity in the vertical length scale can only be shown in the special case when $\theta=0^\circ$, where the base flow is not imposing another length scale to the flow. When the cylinder is tilted (i.e. $\theta \neq 0^\circ$), the geometry of the base flow would externally introduce another vertical length scale \mbox{\citep{Billant2001}} and this argument would not necessarily be valid.}

\begin{table}
  \begin{center}
\def~{\hphantom{0}}
  \begin{tabular}{lcccccccc}
   References & Shear &  $\mathscr{R}(= \Rey\Frou^2)$  & Primary instability  &  Approaches  \\[5pt]
   \citet{Billant2000a,Billant2000c} & H & $O(10)$      & 3D &  EXP/LT \\
   \citet{Deloncle2007} & H &  $\infty$     & 2D/3D  & LT   \\
   \citet{Lucas2017} & H & $O(1-10^2) $  & 2D/3D  &  LT/NT  \\
   \citet{Facchini2018a} & H &  $O(10^2-10^4) $  & 3D  &  EXP/NS/LT  \\
   \citet{Candelier2011} & H/T  &  $\infty$ & 2D*  &  LT \\
   \citet{Meunier2012} & H/T  &  $O(10^{-1}-10) $ & 2D  &  EXP/NS \\
   Present study & H/T  &  $O(10^{-4} - 10) $   & 2D  &  LT with ST
  \end{tabular}
  \caption{\added{A summary of two- and three-dimensional nature of primary instabilities observed in strongly stratified shear flows. Here, the acronyms indicate: H, horizontal; H/T, horizontal and tilted; 2D, two-dimensional; 3D, three-dimensional; EXP, experiment; NS, numerical simulation, LT, linear theory; NT, nonlinear theory; ST, Squire-like theorem. For the instabilities marked as `2D/3D', the most unstable mode remains two dimensional, although the three-dimensional modes have growth rate close to that of two-dimensional one due to the $\beta \Frou$ scaling. For the instabilities marked as `2D*', the nature of three-dimensional modes was not fully explored, despite the potential importance of this mode due to high buoyancy Reynolds number.}}
    \label{tab1}
  \end{center}
\end{table}

\added{In table \ref{tab1}, the variety form of instabilities arising from strongly stratified shear flows is summarised. Here, we note that the work by \citet{Billant2000a,Billant2000c,Billant2000b,Billant2001} is on the so-called `zig-zag' instability which arises from from a vertical columnar vortex pair under strongly stratification. Therefore, in this case, it is not relevant to study two-dimensional horizontal instability. In horizontal shear flows, such as vertical columnar vortex pair \citep{Billant2000c}, Bickley jet \citep{Deloncle2007,Candelier2011}, sinusoidal shear flow \citep{Lucas2017}, it is evident that the primary instability is prone to a three-dimensional mode, which varies vertically, as long as the buoyancy Reynolds number $\mathscr{R}$ is greater than $O(1)$. 
In linear theory, this feature appears through the self-similar scaling of $\omega_i \sim f(\beta \Frou)$ where $\omega_i$ is the growth rate of the instability \citep{Deloncle2007}, and this instability subsequently develops into layered coherent structures in turbulent regime underpinned by non-trivial equilibrium states of the given system \citep{Lucas2017}.} 

\added{Despite the relatively well-established importance of vertically varying structures in stratified shear flows, most of previous linear stability analyses have shown that two-dimensional instability mode is the still most unstable \cite[]{Deloncle2007,Candelier2011,Lucas2017}, except \citet{Facchini2018a} who showed that the primary instability in stratified horizontal Couette flow is three dimensional. Here, it is important to note that, except the Couette flow of \citet{Facchini2018a}, all the two-dimensional primary instabilities, reported by the previous studies and the present one, are inflectional ones, regardless of the value of buoyancy Reynolds number. In the present study, we have shown theoretically that such a two-dimensional instability should be most unstable if buoyancy Reynolds number is sufficiently low and that this behaviour is linked with the self-similar scaling of $\omega_i \sim f(\beta \Rey^{-1/2})$ in this regime. This theoretical result is well supported by the experiment by \citet{Meunier2012}, and also suggests that such a two-dimensional instability may arise as the primary instability in laboratory experiments where the buoyancy Reynolds number is often quite small.}

\added{Compared to the relatively well-studied horizontal shear flows, tilted shear flows have been much less studied. The numerical and experimental results in the present study and \citet{Meunier2012} suggested that two-dimensional instability would still be most unstable if buoyancy Reynolds number is sufficiently low and the shear flow admits an inflectional instability which typically arises at low Reynolds numbers (e.g. vortex shedding). At high buoyancy Reynolds number, \citet{Candelier2011} showed that such a two-dimensional instability is still most unstable for any tilting angles in Bickley jet. However, it is yet clear whether this nature, which appears to arise essentially from the presence of inflectional instability, would extend to the flows without any inflectional instability, such as Couette flow and uniform shear flow. Furthermore, in a tilted shear flows, the presence of tiling angle introduces another vertical length scale \citep{Billant2001}. Therefore, any theoretical foundations established in horizontal shear flows would not necessarily be valid for large tilting angles.}

\section{Results and discussion \label{sec:Result}}

\subsection{Absolute and convective nature of the primary instability\label{subsec:AU}}

Using the numerical solver described in \S\ref{sec:formulation}, we now compute the neutral curve for absolute instability. Given the Squire's theorem shown in \S \ref{subsec:2D}, we will focus on $\beta=0$ in the remaining of the paper. In particular, we will fix $\theta$ and vary $\Frou$ progressively to find the critical Reynolds number for absolute stability $\Rey_c$ for a given set of $\theta$ and $\Frou$. To efficiently find absolute instability, the secant method is used to seek for the pinching point in the complex planes (i.e. $d \omega/d \alpha=0$), which provides the absolute frequency $\omega_0=\omega(\alpha_0)$ where $\alpha_0$ is the absolute streamwise wavenumber.

\begin{figure}
    \centering{}
    \sidesubfloat[]{
        \includegraphics[width=0.45\columnwidth]{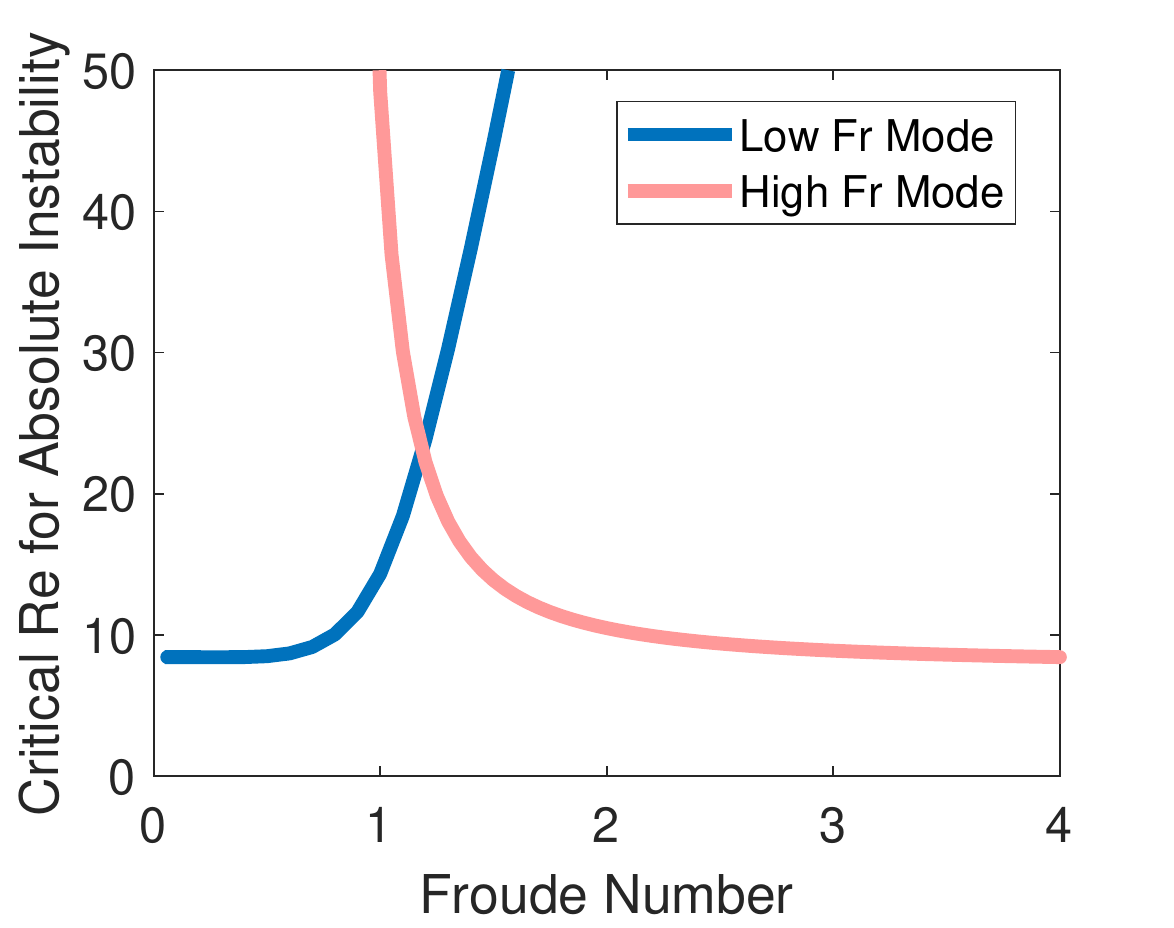}\label{fig:fig5a}
    }
    \sidesubfloat[]{
        \includegraphics[width=0.45\columnwidth]{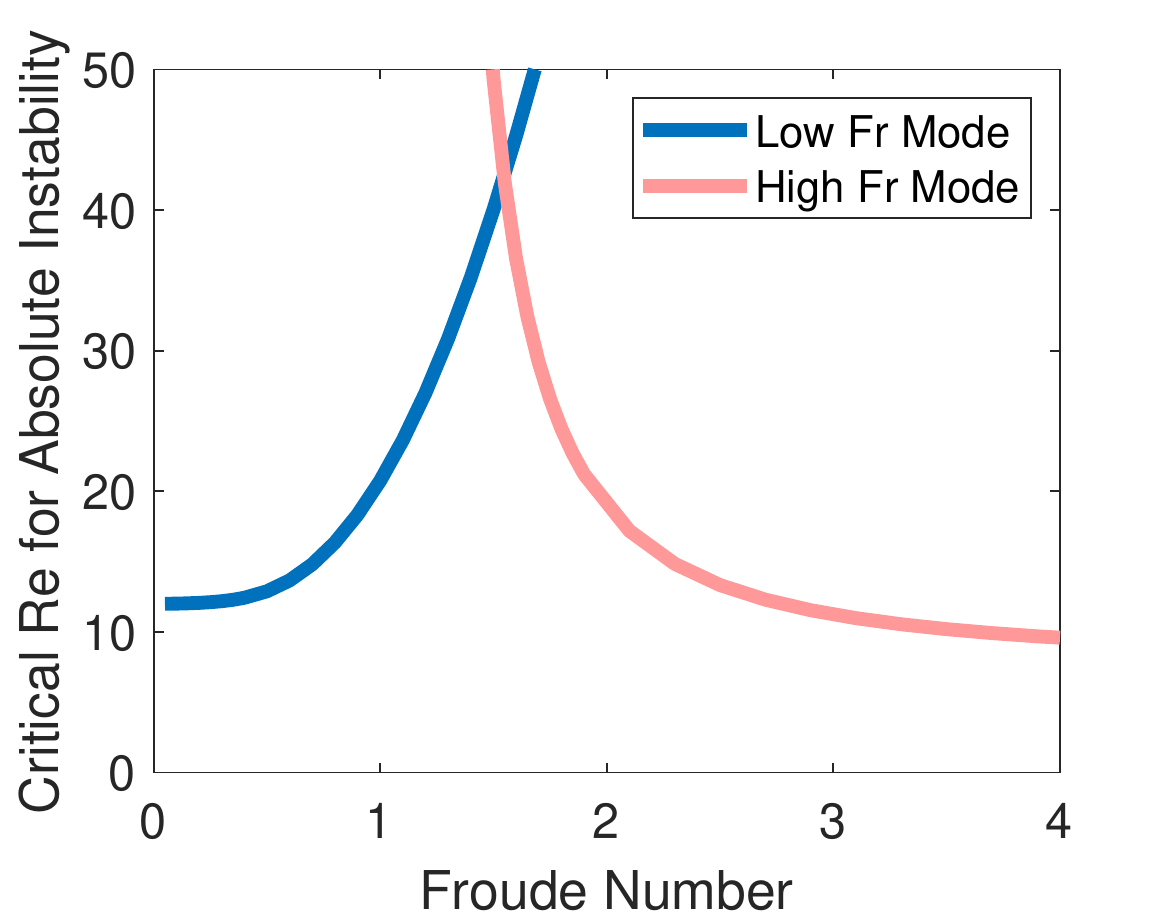}\label{fig:fig5b}
    }
    \caption{The critical Reynolds number for absolute instability with respect to Froude number at $(a)$ $\theta=30^{\circ}$ and $(b)$ $\theta=60^{\circ}$ .\label{fig:fig5}}
\end{figure}

Two modes are found to be most absolutely unstable at a high and a low Froude number respectively, as shown in figure \ref{fig:fig5}. The stability trend is found to be very similar to that of figure 10 in \citet{Meunier2012}. They both show that as the Froude number decreases from the high $\Frou$ regime (say $\Frou>1$), the stratification tends to stabilise the flow. At some low $\Frou$, the low-Froude-number mode becomes important. As its consequence, a sharp cusp emerges in the neutral Reynolds number curve (i.e. the point where the blue (dark) and pink (light) lines meet each other in figure \ref{fig:fig5}), as was also observed in the experimental result \citep{Meunier2012}. With a further decrease of $\Frou$, the flow is more destabilised with a decrease of the critical Reynolds number. This destabilising behaviour with\replaced{ increasing stratification strength (decreasing $\Frou^2$) might sound counter-intuitive,}{ elevation of the stratification level does not sound well intuitively,} and it will be discussed shortly in \S \ref{subsec:finite_Fr}. \added{Lastly, this behaviour of absolute instability for $\beta=0$ is qualitatively the same as that of three-dimensional temporal instability (see Appendix \ref{sec:appen_three}).}

\floatsetup[figure]{style=plain,subcapbesideposition=top}

Despite the good qualitative agreement in the behaviour of the critical Reynolds number with the experimental observation of \citet{Meunier2012}, there is an important difference in the behaviour of $\Rey_c$ with respect to $\theta$. In the present stability analysis, $\Rey_c$ at $\Frou \rightarrow 0$ increases as $\theta$ increases from $30^{\circ}$ to $60^{\circ}$  (blue/darker lines in figure \ref{fig:fig5} at $\Frou \rightarrow 0$). Such a trend was not observed in the experiment of \citet{Meunier2012}, who showed that $\Rey_c$ does not change considerably with such a change of $\theta$. In fact, the behaviour in the present stability analysis is rather similar to that in \citet{Candelier2011}. We will address this issue by proposing some simple explanations in \S \ref{subsec:corrections}.

\begin{figure}
    \centering
    \includegraphics[width=0.7\textwidth]{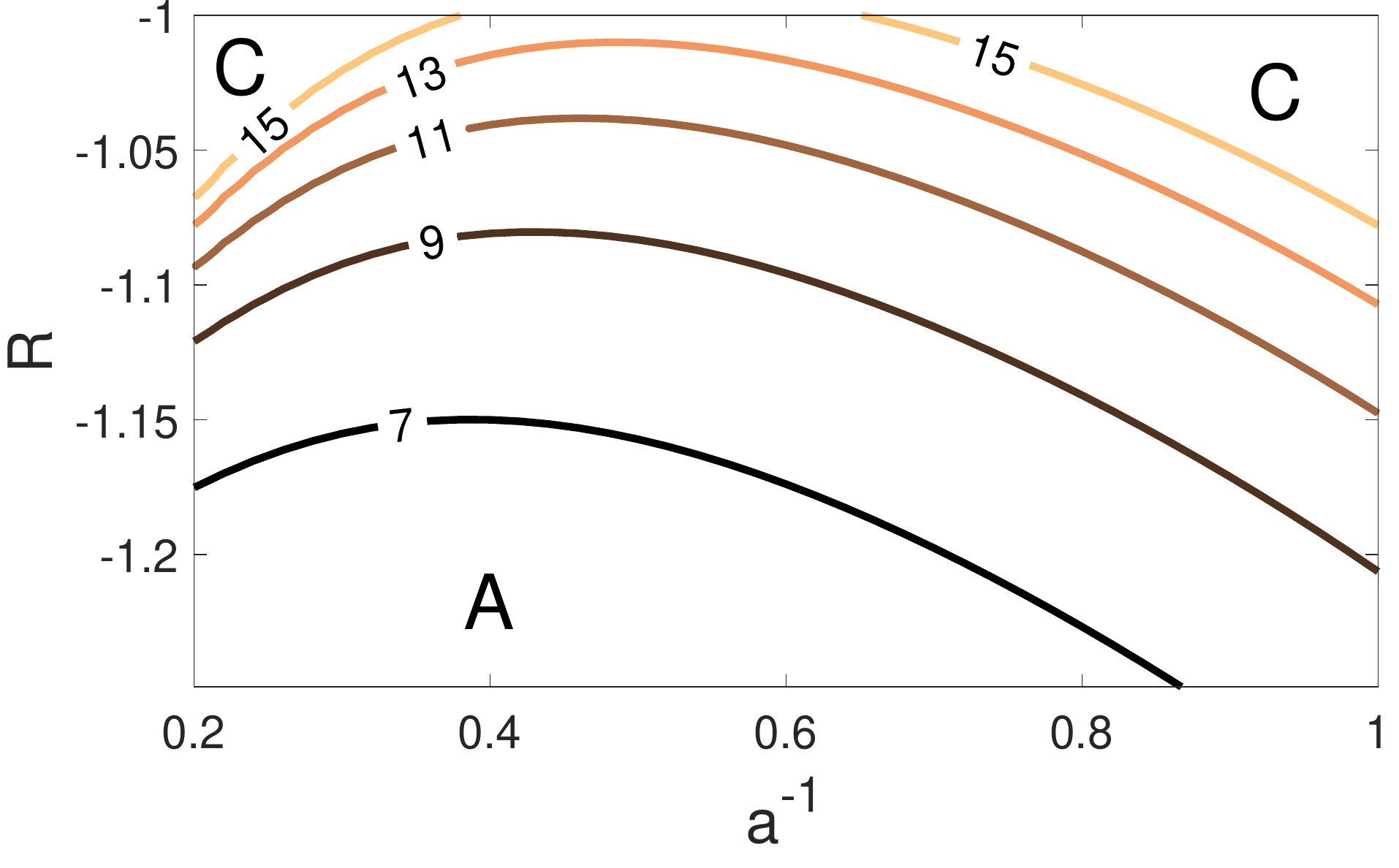}
    \caption{Contour of critical Reynolds number $\Rey_c$ for absolute instability in the $R-a^{-1}$ plane ($\Frou=0.01$ and $\theta=30^{\circ}$). Here, the region marked with `C' is convectively unstable, whereas that with `A' is absolutely unstable. \label{fig:fig6}}
\end{figure}

Finally, to ensure the robustness of the observation made here, the stability analysis for the low-Froude-number mode is repeated by considering a range of base-flow profiles. Figure \ref{fig:fig6} shows contour of the critical Reynolds number obtained by $R$ and $a$ in (\ref{eq:U_profile}). As expected, decreasing $R$ (i.e. having more flow reversal) is found to enhance the absolutely unstable nature of the mode.
For $a$ close to unity, increasing $a$ (i.e. a stiffer profile) destabilises the flow.
The overall behaviour of the critical Reynolds number here is remarkably similar to that found in viscous wakes of homogeneous (i.e. non-stratified) fluid (see figure 4 of \citet{Monkewitz1988}), confirming that the low-Froude-number mode is indeed inflectional and similar to the wake instability in homogeneous fluid. 

\subsection{Stabilisation of the low Froude-number mode with increase of $\Frou$ \label{subsec:finite_Fr}}

In figure \ref{fig:fig5}, we have observed that the low Froude number mode is destabilised as $\Frou$ increases. Since the entire analysis in \S \ref{subsec:Asymptotic-argument} is centred around $W'=0$ at $\Frou \rightarrow 0$ (i.e. (\ref{eq:v_w_same})), here we start our analysis by extending it to the next order.  
We first rewrite (\ref{eq:v_w_same_dev})  in the $(X,Y,Z)$ coordinates: 
\begin{equation}
    W' =\Frou^{2}[\frac{\partial}{\partial t}+U_0 \frac{\partial}{\partial X}-\frac{\nabla^2}{\Rey \Sc}]b', \label{eq:W_scaling}
\end{equation}
indicating that $W'\sim \Frou^2$. If we take $\Frou^2=\epsilon \ll 1$, the following asymptotic expansion can be written:
\begin{subeqnarray}\label{eq:eps_split}
    U'&=& U'_1+\epsilon U'_2 + \textit{O}(\epsilon^2),\\
    V'&=& V'_1+\epsilon V'_2 + \textit{O}(\epsilon^2), \\
    W'&=& W'_1+\epsilon W'_2 + \epsilon^2 W'_3 +\textit{O}(\epsilon^3), \\
    b'&=& b'_1+\epsilon b'_2 + \textit{O}(\epsilon^2), \\
    p'&=& p'_1+\epsilon p'_2 + \textit{O}(\epsilon^2).
\end{subeqnarray}
At $O(\epsilon^{-1})$, (\ref{eq:non_dim_rho}) then yields
\begin{equation}
    W_1'=0,
\end{equation}
retrieving (\ref{eq:v_w_same}). At $\textit{O}(1)$, (\ref{eq:non_dim_u}), (\ref{eq:non_dim_rho}) and the continuity equation in the $(X,Y,Z)$ coordinates can be written as
\begin{subequations}\label{eq:asym_0}
    \begin{eqnarray}
        \left[ \frac{\partial}{\partial t}+U_0 \frac{\partial}{\partial X}-\frac{\nabla^2}{\Rey} \right] U'_1 +  V'_1\frac{dU_0}{dY} & = & -\pardX{p'_1}, \label{eq:asym_u_0} \\
        \left[  \pardt{} +U_0 \pardX{} - \frac{\nabla^2}{\Rey} \right] V'_1  &=& -\pardY{p'_1}  \label{eq:asym_v_0}, \\
        b'_1+\pardZ{p'_1} &=& 0, \label{eq:asym_w_0}\\
        \left[ \frac{\partial}{\partial t}+U_0 \frac{\partial}{\partial X}-\frac{\nabla^2}{\Rey \Sc} \right] b'_1 & = & W'_2, \label{eq:asym_b_0} \\
        \pardX{U'_1}+\pardY{V'_1} & = & 0. \label{eq:asym_cont}
    \end{eqnarray}
\end{subequations}
Here, we note that (\ref{eq:asym_u_0}), (\ref{eq:asym_v_0}) and (\ref{eq:asym_cont}) are decoupled with (\ref{eq:asym_w_0}) and (\ref{eq:asym_b_0}). Furthermore, they can be combined to recover (\ref{eq:v_lowFr_No_Modal}), indicating that this is merely a different derivation of (\ref{eq:v_lowFr_No_Modal}) obtained in the $(X,Y,Z)$ coordinates.
For the same reason, (\ref{eq:asym_w_0}) is also identical to (\ref{eq:non_dim_pb}), and it indicates the hydrostatic balance of $b_1'$ caused by the vertical velocity perturbation at $O(\epsilon)$ (i.e. $W_2'$).

Since (\ref{eq:asym_0}) is identical to the low-Froude-number approximation of (\ref{eq:lineareq}) given in \S\ref{subsec:Asymptotic-argument}, we further proceed to the next order. At $\textit{O}(\epsilon)$, the equations of motion are
\begin{subequations}\label{eq:asym_1}
    \begin{eqnarray}
        \left[ \frac{\partial}{\partial t}+U_0 \frac{\partial}{\partial X}-\frac{\nabla^2}{\Rey} \right] U'_2 + V'_2\frac{dU_0}{dY}  + W'_2\frac{dU_0}{dZ}& = & -\pardX{p'_2}, \label{eq:asym_u_2} \\
        \left[  \pardt{} +U_0 \pardX{} - \frac{\nabla^2}{\Rey} \right] V'_2  &=& -\pardY{p'_2}  \label{eq:asym_v_2}, \\
        \left[  \pardt{} +U_0 \pardX{} - \frac{\nabla^2}{\Rey} \right] W'_2  &=& -\pardZ{p'_2}-b'_2  \label{eq:asym_w_2}, \\
        \pardX{U'_2}+\pardY{V'_2}+ \pardZ{W'_2} &=& 0, \label{eq:asym_cont1}\\
        \left[ \frac{\partial}{\partial t}+U_0 \frac{\partial}{\partial X}-\frac{\nabla^2}{\Rey \Sc} \right] b'_2 & = & W'_3 . \label{eq:asym_b_2}
    \end{eqnarray}
\end{subequations}
Now, it becomes evident that the key structural difference between (\ref{eq:asym_0}) and (\ref{eq:asym_1}) is the presence of $W'_2$ in (\ref{eq:asym_1}) -- indeed, if $W'_2=0$, the form of (\ref{eq:asym_0}) is identical to that of (\ref{eq:asym_1}). This implies that the presence of the non-zero vertical velocity (i.e. $W_2'$) is the key player for the stabilisation mechanism of the low-Froude-number mode on increasing $\Frou$ from a very small value. Furthermore, the form of (\ref{eq:asym_1}) suggests that there may be two possible stabilisation mechanisms played by $W'_2$: one is modification of the shear instability through an interaction with the vertical shear (i.e. $W'_2 (dU_0/dZ)$ in (\ref{eq:asym_u_2})), and the other is coupling with the stabilising buoyancy through (\ref{eq:asym_w_2}). Despite the useful physical insight gained here, it is difficult to solve (\ref{eq:asym_1}) even numerically. This is because of the unknown $W_3'$ in (\ref{eq:asym_b_2}), which will have to be obtained from the equations at $O(\epsilon^2)$. Unfortunately, the form of the equations at $O(\epsilon^2)$ is exactly identical to that of (\ref{eq:asym_1}), requiring the vertical velocity perturbation at $O(\epsilon^3)$. In fact, this pattern repeats in the equations at any subsequent orders, creating a closure problem that makes it difficult to proceed the current analysis any further. 

\subsection{Energy budget analysis \label{subsec:budget}}

\floatsetup[figure]{style=plain,subcapbesideposition=top}
\begin{figure}
    \centering
    \sidesubfloat[]{
        \includegraphics[width=0.42\textwidth]{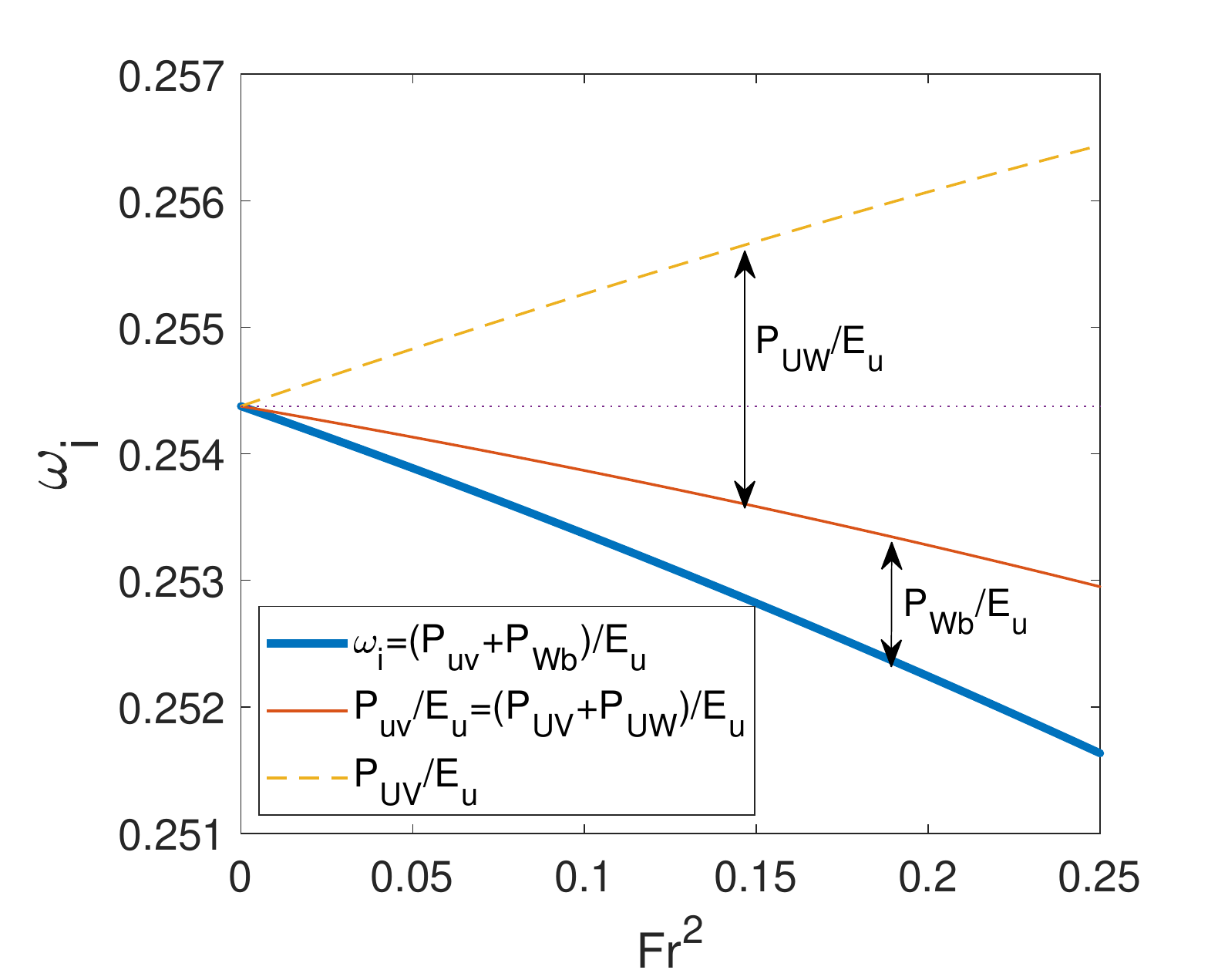}
        \label{fig:fig7a}
    }
    \sidesubfloat[]{
        \includegraphics[width=0.42\textwidth]{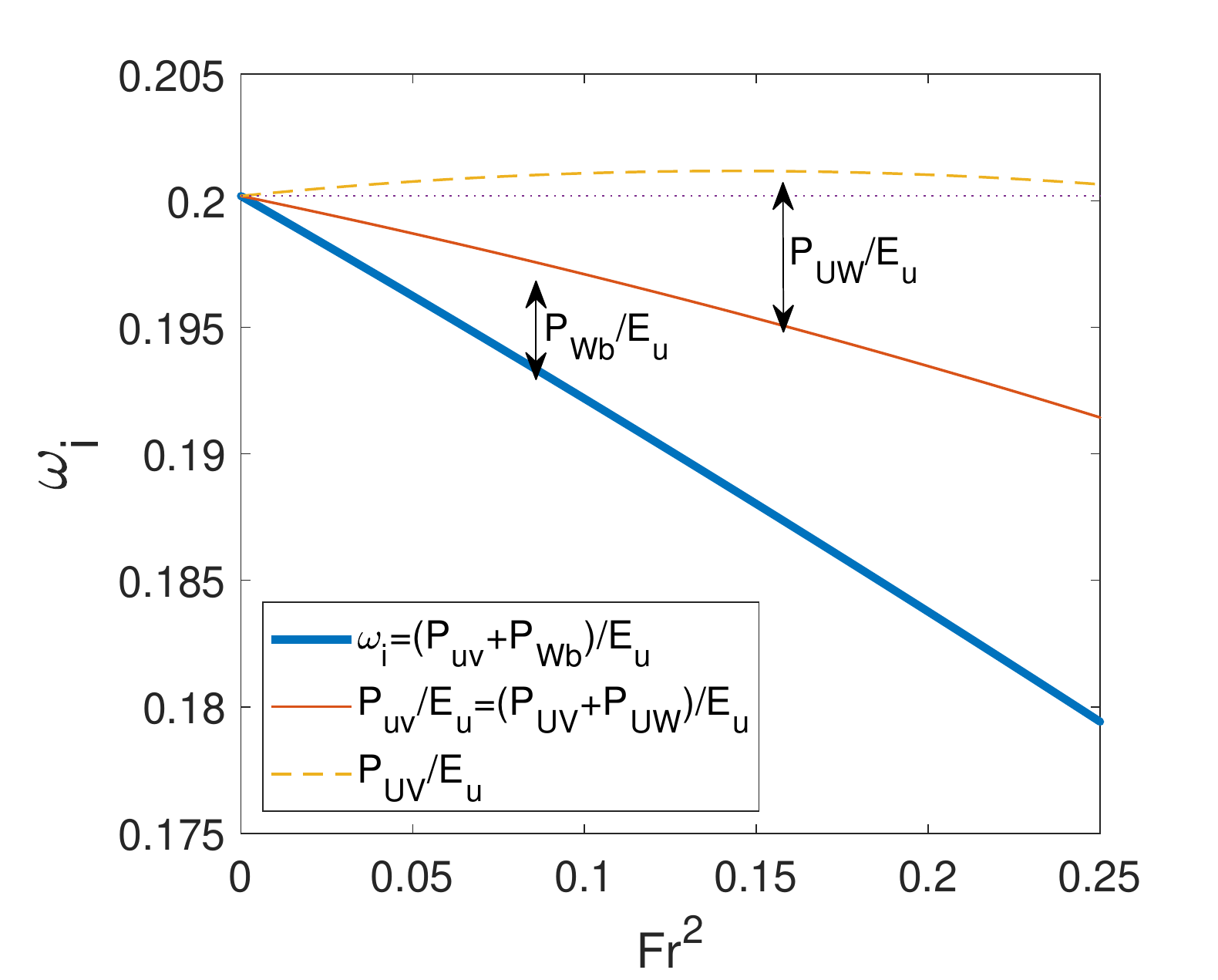}
        \label{fig:fig7b}
    }
    \caption{Contribution of production ($P_{uv}/E_u$) and conversion to potential energy per kinetic energy to the growth rate $\omega_i$: $(a)$ $\theta=30^\circ, \alpha=0.4$; $(b)$ $\theta=60^\circ, \alpha=0.4$. Here, note that the difference between the (blue) thick line ($\omega_i$) and (red) thin line ($P_{UV}+P_{UW}/E_u$) indicates the contribution of $P_{Wb}/E_u$, while the difference between the (red) thin line and the dashed lined ($P_{UV}/E_u$) indicates the contribution of $P_{UW}/E_u$. For reference, the horizontal dotted line shows the growth rate at $\Frou=0.01$. \label{fig:fig7}}
\end{figure}

Given the difficulty discussed in \S\ref{subsec:finite_Fr}, we now proceed to explore the more precise stabilisation mechanisms using the numerical result. In particular, we perform an energy budget analysis of the perturbation state to understand the role of two mechanisms discussed above. For simplicity, we shall now consider only the inviscid case. 

We introduce total energy of the perturbed state in the eigenvalue problem (\ref{eq:normalmode}), such that
    \begin{subequations}
    \begin{equation}
        E_{tot}=E_{u}+E_{b},    
    \end{equation}
with
    \begin{equation}
        E_{u}=\int_{-\infty}^{\infty} |\tilde{u}|^2 + |\tilde{v}|^2 + |\tilde{w}|^2 ~dy,~~E_{b}= \int_{-\infty}^{\infty}\Frou^2|\tilde{b}|^2~dy, \label{eq:Full_5_energy_E}
    \end{equation}
\end{subequations}
where $E_u$ is the kinetic energy of the perturbed state and $E_b$ is the potential energy. The energy balance of (\ref{eq:normalmode}) is given by 
\begin{subequations}\label{eq:Full_5_energy_total}
    \begin{equation}
        \omega_{i} (E_{u}+E_{b})=P_{uv} ,
    \end{equation}
with
    \begin{equation}
        P_{uv}=-\int_{-\infty}^{\infty}\Real(\bar{\tilde{u}}\tilde{v}DU)~dy,
    \end{equation}
\end{subequations}
where $\bar{\cdot}$ indicates complex conjugate (see Appendix for details). Equation (\ref{eq:Full_5_energy_total}) now indicates that $P_{uv}$ (i.e. production by the given base flow) is the only source term of the instability. The contribution of the production $P_{uv}$ can be split into $E_u$ and $E_{b}$, the former of which becomes kinetic energy of the instability mode and the latter is potential energy converted by buoyancy. Furthermore, from (\ref{eq:lin_rho}), the potential energy balance is written as
\begin{subequations}
    \begin{equation}
        \omega_i E_{b}=-P_{vb}-P_{wb},
    \end{equation}
with
    \begin{equation}
        P_{vb}=\int_{-\infty}^{\infty}\Real(\bar{\tilde{v}}\sin\theta\tilde{b})dy,~~
        P_{wb}=-\int_{-\infty}^{\infty}\Real(\bar{\tilde{w}}\cos\theta\tilde{b}) dy,
    \end{equation}
\end{subequations}
where $P_{vb}$ and $P_{wb}$ are the rate of potential energy generation by $v'$ and $w'$, respectively. Then, the kinetic energy balance of the instability mode becomes
\begin{equation}
    \omega_{i} E_{u}=P_{uv}+P_{wb}+P_{vb}. \label{eq:Full_5_energy_sim_nonzero}
\end{equation}

Given the importance of the vertical velocity perturbation discussed in \S\ref{subsec:finite_Fr}, we now convert (\ref{eq:Full_5_energy_sim_nonzero}) into the one in the $(X,Y,Z)$ coordinates. Using $\tilde{U}(y)=\tilde{u}$, $\tilde{V}(y)=\tilde{v}\cos \theta +\tilde{w}\sin \theta$ and  $\tilde{W}(y)=\tilde{w}\cos \theta -\tilde{v}\sin \theta$, the kinetic energy balance in the $(X,Y,Z)$ coordinates is given by
\begin{subequations}\label{eq:Full_5_energy_sim_nonzero_lab}
    \begin{equation}
        \omega_{i} E_{U}=P_{UV}+P_{UW}+P_{Wb}, 
    \end{equation}
where
    \begin{equation}
        E_{U}=\int_{-\infty}^\infty  |\tilde{U}|^2 + |\tilde{V}|^2 + |\tilde{W}|^2 dy, \label{eq:Full_5_energy_E_UWV}
    \end{equation}
    \begin{equation}
        P_{UV}= - \int_{-\infty}^\infty \Real \left(\bar{\tilde{U}} \frac{dU_0}{dY} \tilde{V} \right)dy, \quad P_{UW}= - \int_{-\infty}^\infty \Real \left( \bar{\tilde{U}}  \frac{dU_0}{dZ} \tilde{W} \right) dy,
    \end{equation}
    \begin{equation}
        P_{Wb}= - \int_{-\infty}^\infty \Real \left( \bar{\tilde{W}} \tilde{b} \right) dy.
    \end{equation}
\end{subequations}
Here, $E_U=E_u$, $P_{uv}=P_{UV}+P_{WV}$ and $P_{Wb}=P_{vb}+P_{wb}$.
From (\ref{eq:Full_5_energy_sim_nonzero_lab}), it becomes clear that kinetic energy of the instability mode is formed by balance of two mechanisms: 1) production $P_{UV}+P_{UW}$ by the given base-flow shear in the horizontal and vertical directions; 2) conversion to potential energy (or loss of kinetic energy) through $P_{Wb}$.

Figure \ref{fig:fig7} shows the numerical results of how the production per given kinetic energy ($P_{uv}/E_u$) and the conversion to potential energy per given kinetic energy ($P_{Wb}/E_u$) contribute to the stabilising effect on the low Froude number mode on increasing $\Frou$. As expected, $P_{Wb}$ is stabilising as $\Frou^2$ increases. However, it is also found that the production $P_{uv}$ also significantly decreases, as $\Frou$ is increased. In particular, $P_{UW}$ takes energy from the instability (i.e. stabilising) on increasing $\Frou$, while $P_{UV}$ plays a destabilising role of the flow. 

\begin{figure}
    \centering
    \includegraphics[width=0.98\textwidth]{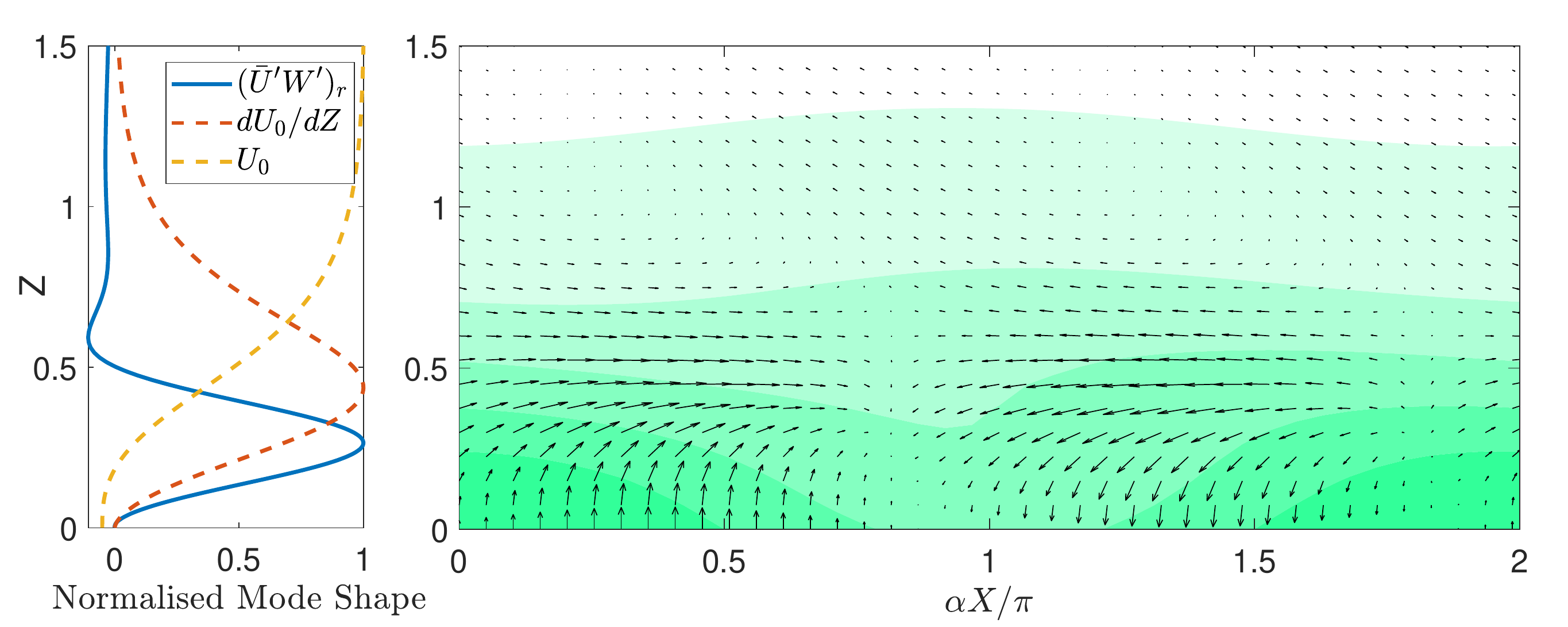}
    \caption{On the left shows the normalised value of $\bar{U}'W'$ (real) of the most unstable mode, the base flow and its derivatives in $Z$ projected on $Z$ axis, where each plot is normalised by its own maximum value. On the right shows the spatial structure of $U'$ and $W'$ of the most unstable mode on the $X-Z$ plane, and the contour plot of density perturbation $b'$ added to the linear and stable background stratification. In this figure, $\alpha=0.4$, $\Frou=0.5$ and $\theta=30^\circ$. Note that $W'$ is scaled up by a arbitrary value for visualisation purpose. \label{fig:fig8} }
\end{figure}

To understand more precise physical picture on the stabilisation mechanisms, the spatial mode structure of $U'$ and $W'$ in the $X$-$Z$ plane is shown for a relatively small $\Frou=0.5$ in figure \ref{fig:fig8}. As expected, the finite $\Frou$ allows for vertical velocity fluctuation $W'$. The positive and negative $W'$ are well correlated to high and low buoyancy fluctuations $b'$, indicating transport of buoyancy field by $W'$ (($X,Z)\simeq (0.3\pi/\alpha,~0.2)$ and ($X,Z)\simeq (0.8\pi/\alpha,0.2)$) in figure \ref{fig:fig8}): i.e. the high $b'$ is transported from the lower region where basic-state buoyancy is high, and the high $b'$ is transported from the upper region with low buoyancy field. It is evident that such high and low buoyancy fluctuations would create upward and downward gravitational force, directions of which are opposite to those of $W'$. Therefore, this mechanism would play a stabilising role, consistent with the action of $P_{Wb}$ in (\ref{eq:Full_5_energy_sim_nonzero_lab}). It is also interesting to observe the positive and negative $W'$ emerge in the regions where the streamwise velocity fluctuations are respectively positive and negative (($X\simeq 0.25\pi/\alpha,~Z=0.2$) and ($X\simeq 1.25\pi/\alpha,~Z=0.2$) in figure \ref{fig:fig8}). This suggests that the low-Froude-number inflectional instability mode interacts with $W'$, such that a positive correlation between $U'$ and $W'$ is generated, providing a stabilising mechanism through $P_{UW}$ in (\ref{eq:Full_5_energy_sim_nonzero_lab}).  

\subsection{Simple corrections of base flow for consistent comparison with experimental data \label{subsec:corrections}}

\floatsetup[figure]{style=plain,subcapbesideposition=top}
\begin{figure}
    \sidesubfloat[]{
        \includegraphics[width=0.42\columnwidth]{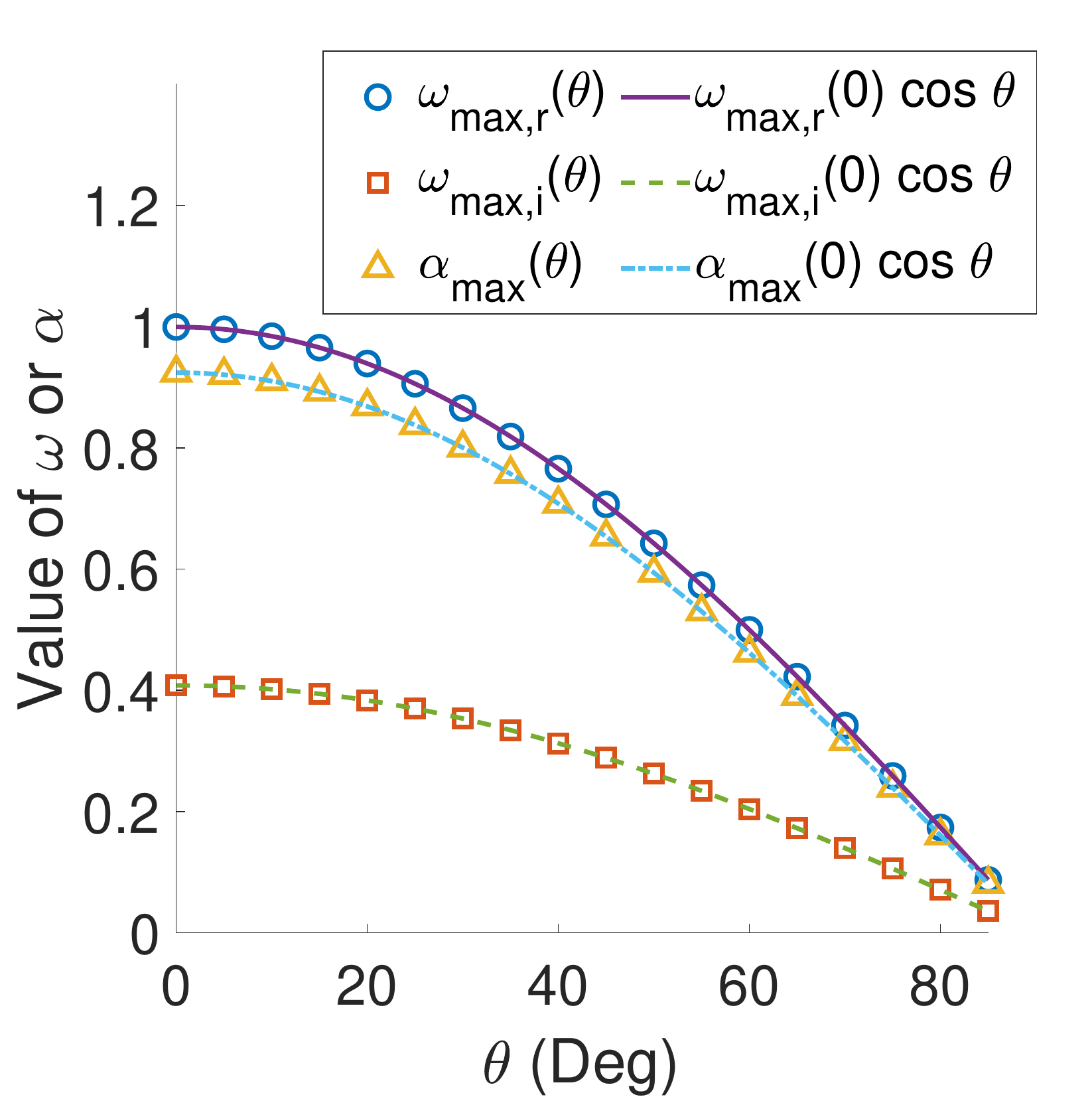}
        \label{fig:fig9a}
    }
    \sidesubfloat[]{
        \centering{}
        \includegraphics[width=0.42\columnwidth]{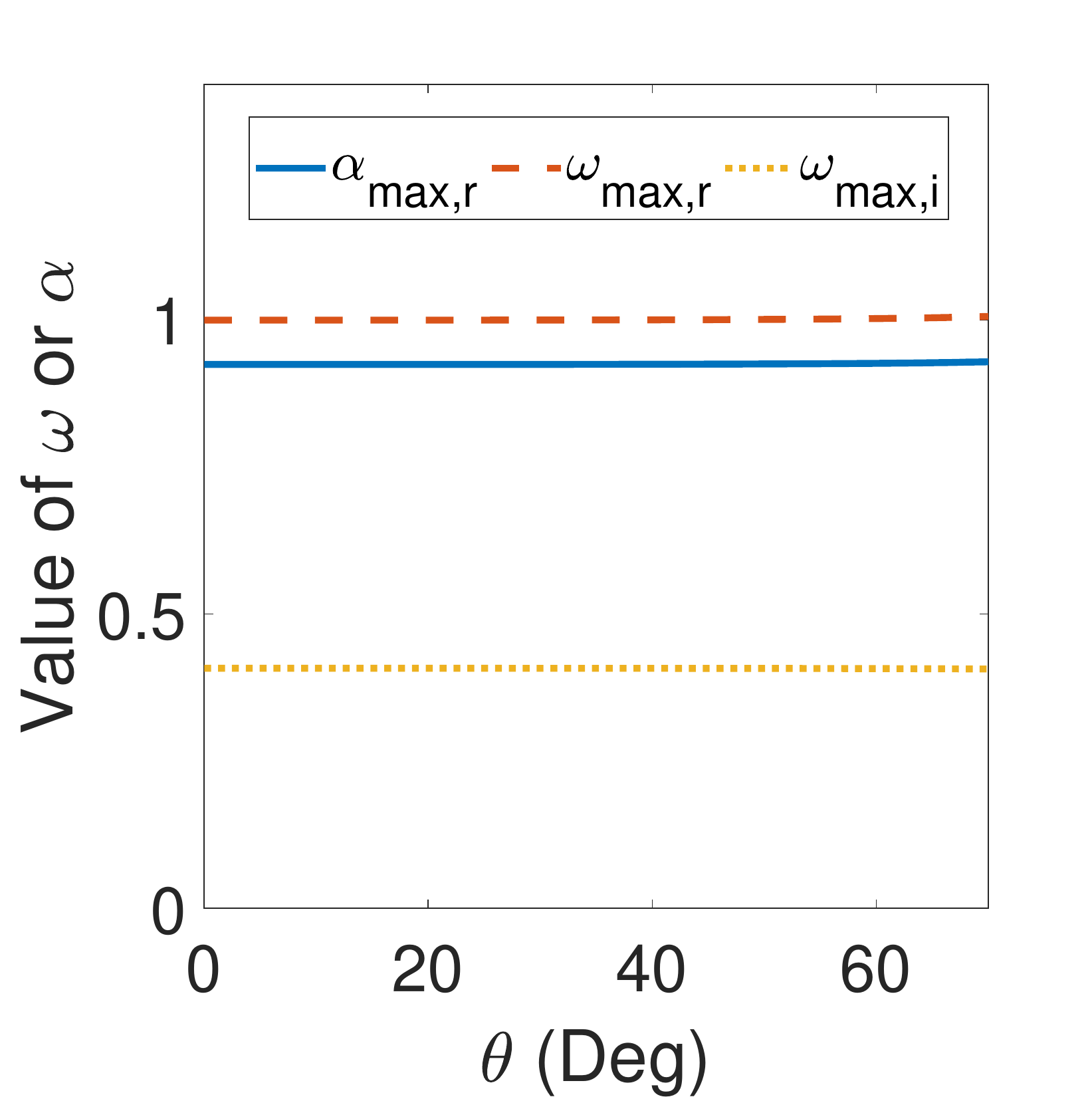}
        \label{fig:fig9b}
    }
    \caption{The most unstable frequency $\omega_{max}$ (complex) and streamwise wavenumber $\alpha_{max}$ (real) of the temporal instability of an inviscid wake as a function of $\theta$ ($\Frou=0.01$) with $(a)$ a fixed base-flow profile $u_0(y)$ on the $x$-$y$ plane and $(b)$ a fixed base-flow profile $U_0(Y)$ on the $X$-$Y$ plane. Note that, in $(a)$, the fit of $\cos \theta$ curves overlaps with the corresponding numerical data.}
\end{figure}

As discussed in \S\ref{sec:Introduction}, \citet{Candelier2011} showed that the inviscid growth rate of the low-Froude-number instability mode is highly dependant on $\theta$, whereas the experimental result in wake flow by \citet{Meunier2012} showed that $\Rey_c$ remains at a similar level when $\theta=30^\circ$ and $\theta=60^\circ$. Given the nature of the base flow  considered so far (i.e. the base flow kept to be the same in the $(x,y,z)$ coordinates like in \citet{Candelier2011}), it is not surprising to see that the critical Reynolds number for absolute instability in \S \ref{sec:Result} is found to considerably vary with $\theta$. Therefore, in the remaining part of this paper, we will explore what physical processes would need to be further modelled in order to make a consistent comparison between the present stability analysis and the experimental data in \citet{Meunier2012}. 

Here, we propose three possible physical origins that may explain the aforementioned difference between the present study and \citet{Meunier2012}: 1) the nature of the base flow kept to be the same in the $(x,y,z)$ coordinates; 2) the effect of viscosity; 3) the effect of changing cross-section of the given bluff body on the horizontal plane with $\theta$ in the experiment of \citet{Meunier2012}. It should be stressed that the propositions given here are based on physical understanding of the flow and comparison with experiments. Therefore, they should be viewed as observation-based suggestions that may help to improve modelling of `real' base flow with the change of the tilting angle $\theta$ in the experiment of \citet{Meunier2012}.

\subsubsection{Base flow on the horizontal plane \label{subsec:Base-Flow}}

According to \citet{Candelier2011}, the stability of an inviscid tilted shear flow under the low-Froude-number approximation ($\Frou \rightarrow 0$) satisfies the following relationship:
\begin{equation}
    \omega(\alpha,\Frou,\theta)=\omega(\alpha/\cos\theta,0,0)\cos\theta, \label{eq:inviscid_scaling_law}
\end{equation}
given the base flow is kept unchanged on the tilted plane (i.e. $x$-$y$ plane). This dispersion relation can also be demonstrated numerically in the present case by taking the inviscid limit, as shown in figure \ref{fig:fig9a}.

Here, the dispersion relation (\ref{eq:inviscid_scaling_law}) can be interpreted as the stability of a tilted base flow being the same as that of its projection on the horizontal plane. Indeed, if the base flow is kept the same on the tilted plane, the width (i.e. the length scale of the system) of the base flow projected on the horizontal plane is rescaled by a factor of $(\cos \theta)^{-1}$. This is mathematically equivalent to multiplying $Y$ in (\ref{eq:v_lowFr_No_Modal}) by this factor recovering (\ref{eq:inviscid_scaling_law}) from (\ref{eq:v_lowFr_No_Modal}) in the inviscid limit. Therefore, if the base-flow projection on the horizontal plane $U_0(Y)$ is enforced to be unchanged with respect to $\theta$ (instead of a constant base flow $u_0(y)$ on the tilted plane), the inviscid stability at $\Frou\rightarrow0$ should be independent of $\theta$. This is also demonstrated numerically in figure \ref{fig:fig9b}.

\subsubsection{Viscosity and bluff-body geometry \label{subsec:ellipse_effect}}

The independence of $\omega$ on $\theta$ obtained by the use of unchanging base flow on the horizontal plane with $\theta$ provides an important explanation on the discrepancy between the results of the stability analysis and the experiment. 
However, in the low Reynolds number regime where the experiment by \citet{Meunier2012} was performed, one should not ignore the effect of viscosity: given the strong vertical stratification, $\partial^2/\partial Z^2$ in the viscous dissipation term in (\ref{eq:v_lowFr_No_Modal}) can certainly introduce $\theta$-dependence. 

\begin{figure}
    \centering{}
    \sidesubfloat[]{
        \includegraphics[width=0.45\columnwidth]{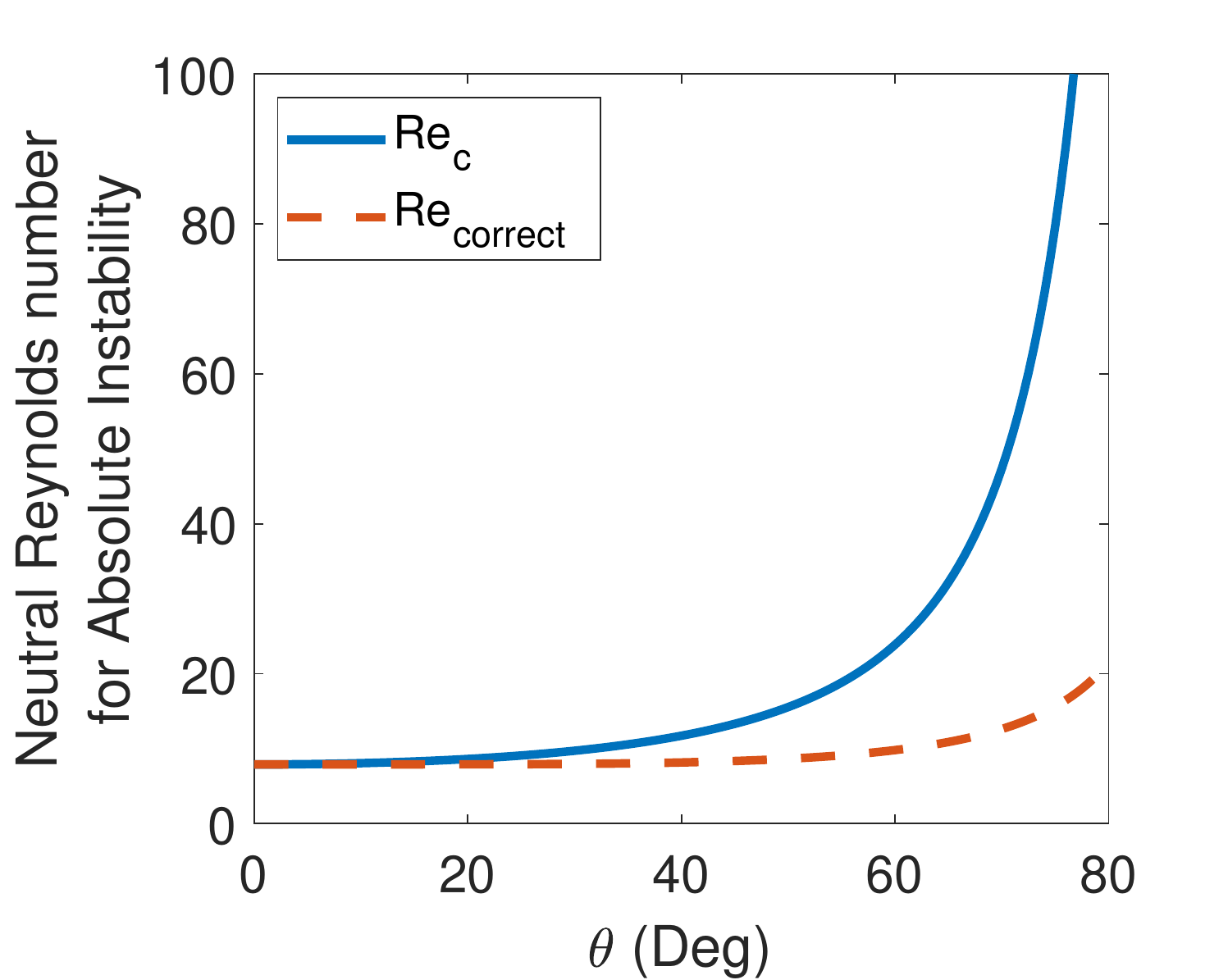}
        \label{fig:fig10a}
    }
    \sidesubfloat[]{
        \includegraphics[width=0.45\columnwidth]{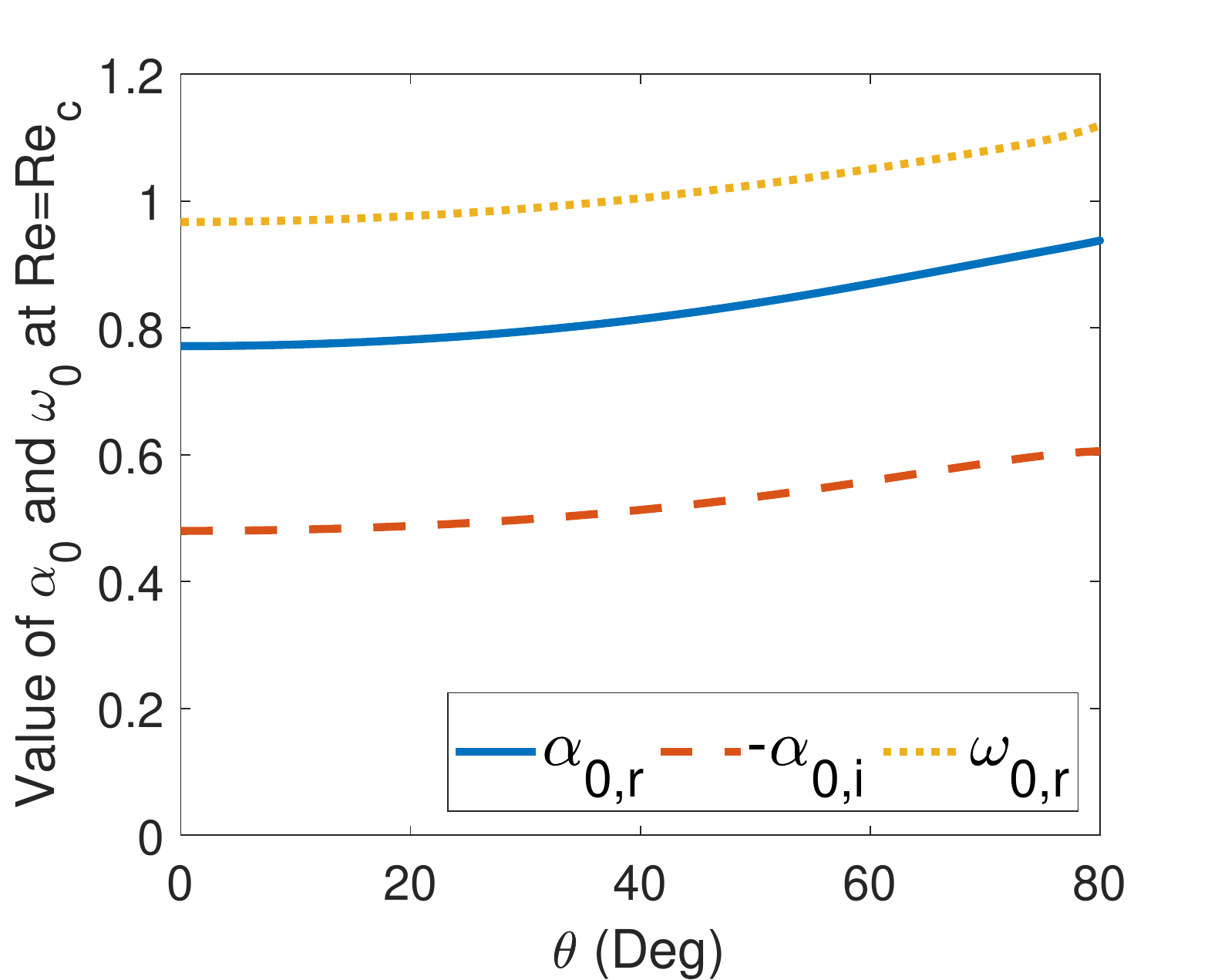}
        \label{fig:fig10b}
    }
    \caption{The behaviour of neutral absolute instability mode ($\Frou=0.01$): $(a)$ $\Rey_c$; $(b)$ $\omega_{0}$ and $\alpha_{0}$. The base-flow profile on the horizontal plane is kept to be the same as $\theta$ changes (see also text).}
\end{figure}

We have re-computed the viscous absolute instability at $\Frou = 0.01$ ($\Frou \rightarrow 0$) with an unchanging $U_0(Y)$ profile. It is found that critical Reynolds number $\Rey_c$ is still dependent on the tilting angle $\theta$ in the viscous case, as shown in figure \ref{fig:fig10a} (blue line). Figure \ref{fig:fig10b} shows that the corresponding absolute streamwise wavenumber $\alpha_0$ does not change much like in figure \ref{fig:fig9b}. This suggests that the increase of the critical Reynolds number is likely to be caused by the viscous term in (\ref{eq:v_lowFr_No_Modal}).

\floatsetup[figure]{style=plain,subcapbesideposition=top}

Given the effect of viscosity on the critical Reynolds number, it appears that keeping base flow the same on the horizontal plane for different $\theta$ would not fully explain the different $\theta$-dependency of the instability between the stability analysis and the experiment. Therefore, we should finally consider the effect of the changing horizontal cross-sectional geometry with the tilting angle $\theta$, which would also be significant in a real bluff-body wake. It is evident that increasing tilting angle would result in a more elongated horizontal cross-sectional body (an ellipse stretched in the $Y$ direction in the case of circular cylinder). As such, we hypothesise that the more elongated the horizontal section of the given body is, the more flow is destabilised. In other words, the more elongated horizontal cross section in the $Y$ direction would increase the length scale of the given system, thereby increasing the effective Reynolds of the system. This may then reduce $\Rey_c$ at higher $\theta$. 

Now, we will check this hypothesis by empirically correcting $\Rey_c$ from the viscous analysis with the formula
\begin{equation}\label{eq:corr}
    \Rey_{correct}=\Rey_c C_{shape}(\theta)\cos\theta  ,
\end{equation}
where $C_{shape}$ is the correction factor for the elongated ellipse.
Such a factor is assumed to be the ratio between $\Rey_{c}$ of an elliptical cylinder and $\Rey_{c}$ of a circular cylinder, which can be empirically calculated from the data \citet{Thompson2014}. We note that, in (\ref{eq:corr}), the factor $\cos \theta$ is introduced, so that the length scale used in \citet{Thompson2014} becomes consistent with that in the present study: \citet{Thompson2014} kept the width of the ellipse unchanged while shortening the streamwise length, whereas, in our analysis, the streamwise length is kept unchanged and the width is set to be elongated.

The corrected neutral Reynolds number $\Rey_{correct}$ is shown in figure \ref{fig:fig10a} (red dashed line), together with the uncorrected neutral Reynolds number $\Rey_c$ (blue solid line). The behaviour of $\Rey_{correct}$ now appears to be more consistent with that in \citet{Meunier2012}, as $\Rey_{correct}$ at $\theta=30^{\circ}$ and $\theta=60^{\circ}$ are of similar values. While one may argue that the corrected result has qualitatively reconciled with the data from \citet{Meunier2012}, we should admit that this still remains a proposition that needs to be confirmed with experiments or simulations. In particular, a further examination appears to be required on how reliable such an ad-hoc correction is. The empirical law proposed by \citet{Meunier2012} suggested that $\Rey_c$ is not a strong function of $\theta$ at $\Frou \rightarrow 0$, but the corrected result is still suggesting a stabilising effect at higher $\theta$.

\section{Conclusion \label{sec:conclusion}} 

This paper aims to gain a better understanding of the low-Froude-number mode using a linear stability analysis.
It has successfully reproduced the stability result of the experiment of \citet{Meunier2012}, in which there exists a branch switching from the high-Froude-number mode to the low-Froude-number one when $\Frou$ decreases.

To better understand the nature of the low-Froude-number mode, we have put forward the use of laboratory frame (i.e. the $(X, Y, Z)$ coordinate system). Using the laboratory frame and the approximation at $\Frou \rightarrow 0$, we have simplified the set of linearised equations of motion to (\ref{eq:v_lowFr_No_Modal}), which emerges in a very similar form of the Orr-Sommerfeld equation in physical space. Based on (\ref{eq:v_lowFr_No_Modal}), we have deduced that the low-Froude-number mode observed in  \citet{Meunier2012} and in the present stability analysis is presumably a two-dimensional and horizontal (barotropic) inflectional instability.

Using (\ref{eq:v_lowFr_No_Modal}) and the WKBJ approximation, we have shown that the most unstable mode is indeed two-dimensional as long as the tilting is weak\added{, while the numerical result extends the two-dimensional argument to all $\theta$ at low $\Frou$}. The physical understanding of this theoretical result is that any vertical variation in the small perturbation would only introduce more viscous dissipation, thereby stabilising the given instability. It is important to mention that, at $\theta=0^\circ$, this result is valid only in the regime of low buoyancy Reynolds number $\mathscr{R}=\Rey \Frou$ \citep[see also][]{Brethouwer2007} where the self-similarity is observed with respect to $\beta Re^{-1/2}$.

We also investigated how the increasing $\Frou$ from $\Frou=0$ stabilise the system, especially given our understanding that the low Froude number mode is horizontal and two-dimensional at $\Frou=0$. Using an asymptotic expansion and energy budget analysis, we have observed that the emergence of small vertical velocity at $\textit{O}(\Frou^2)$ plays the key role in stabilisation of the flow on increasing $\Frou^2$. This stabilisation mechanism is also found to be associated with the paradoxically stabilising buoyancy on increasing Froude number and with the modification of inflectional instability.

We have also tried to explain the different $\theta$-dependency of the low-Froude-number instability mode of \citet{Candelier2011} from \citet{Meunier2012} by proposing a suitable behaviour of base flow with respect to the tilting angle. While the proposed base-flow corrections may yield a reasonable agreement between the stability analysis and the experiment, they still remain to be tested. \added{In this respect, a global stability analysis with a base flow obtained from a full numerical simulation would be highly desirable to confirm the propositions we have made in the present study. This would be the important next step towards complete understanding of instabilities in titled stratified bluff-body wake.} 


\section*{Acknowledgement}
L. F. gratefully acknowledges funding from the President's PhD Scholarship of Imperial College London. We would also like to thank Professor C. \replaced{P.}{C.} Caulfield and Dr P. Billant for the insightful discussions. L. F. is grateful to Dr P. Meunier who shared his experience with the experiment.

\appendix
\section{Derivation of energy budget analysis \label{sec:appen_energy}}

Now, (\ref{eq:lin_u})-(\ref{eq:lin_incompress}) can be rewritten as:

\begin{equation}
    i\omega E_{total}= \int_{-\infty}^{\infty}
    \left(\begin{array}{c}
        \bar{\tilde{u}}\\
        \bar{\tilde{v}}\\
        \bar{\tilde{w}}\\
        \bar{\tilde{b}}\\
        \bar{\tilde{p}}
    \end{array}\right)^T
    \left(\begin{array}{ccccc}
        \mathcal{L} & DU & 0 & 0 & i\alpha\\
        0 & \mathcal{L} & 0 & -\sin\theta & D\\
        0 & 0 & \mathcal{L} & \cos\theta & 0\\
        0 & \sin\theta & -\cos\theta & \mathcal{L}_{\rho} \Frou^2 & 0\\
        i\alpha & D & 0 & 0 & 0
    \end{array}\right)
    \left(\begin{array}{c}
        \tilde{u}\\
        \tilde{v}\\
        \tilde{w}\\
        \tilde{b}\\
        \tilde{p}
    \end{array}\right) dy,
\end{equation}
where
\begin{equation}
    E_{total}=\int_{-\infty}^{\infty} |\tilde{u}|^2 + |\tilde{v}|^2 + |\tilde{w}|^2 + \Frou^2 |\tilde{b}|^2 dy. \label{eq:Full_5_energy_E_app}
\end{equation}

We note that we have deliberately scaled  (\ref{eq:lin_rho}) by $\Frou^2$ to recover an energy budget that has physical significance. Therefore, the first three terms in the integrand of (\ref{eq:Full_5_energy_E_app}) form
\begin{equation}
    E_{u}=\int_{-\infty}^{\infty} |\tilde{u}|^2 + |\tilde{v}|^2 + |\tilde{w}|^2 dy,
\end{equation}
representing the kinetic energy, and the last term
\begin{equation}
    E_{b}=\int_{-\infty}^{\infty} \Frou^2 |\tilde{b}|^2 dy
\end{equation}
being the potential energy.

Now, since we are only interested in the growth rate, the imaginary part of $\omega$, we can therefore define the following terms:

\begin{subeqnarray}
    P_{uu}= & -\Real(\int_{-\infty}^{\infty}\bar{\tilde{u}}\mathcal{L}\tilde{u}dy),\\
    P_{uv}= & -\Real(\int_{-\infty}^{\infty}\bar{\tilde{u}}DU\tilde{v}dy),\\
    P_{up}= & -\Real(i\alpha\int_{-\infty}^{\infty}\bar{\tilde{u}}\tilde{p}dy),\\
    P_{vv}= & -\Real(\int_{-\infty}^{\infty}\bar{\tilde{v}}\mathcal{L}\tilde{v}dy),\\
    P_{vb}= & -\Real(-\sin\theta\int_{-\infty}^{\infty}\bar{\tilde{v}}\tilde{b}dy),\\
    P_{vp}= & -\Real(\int_{-\infty}^{\infty}\bar{\tilde{v}}D\tilde{p}dy),\\
    P_{ww}= & -\Real(\int_{-\infty}^{\infty}\bar{\tilde{w}}\mathcal{L}\tilde{w}dy),\\
    P_{wb}= & -\Real(\cos\theta\int_{-\infty}^{\infty}\bar{\tilde{w}}\tilde{b}dy),\\
    P_{bv}= & -\Real(\sin\theta\int_{-\infty}^{\infty}\bar{\tilde{b}}\tilde{v}dy),\\
    P_{bw}= & -\Real(-\cos\theta\int_{-\infty}^{\infty}\bar{\tilde{b}}\tilde{w}dy),\\
    P_{bb}= & -\Real(\int_{-\infty}^{\infty}\Frou^2 \bar{\tilde{b}}\mathcal{L}_{\rho}\tilde{b}dy),\\
    P_{pu}= & -\Real(i\alpha\int_{-\infty}^{\infty}\bar{\tilde{p}}\tilde{u}dy),\\
    P_{pv}= & -\Real(\int_{-\infty}^{\infty}\bar{\tilde{p}}D\tilde{v}dy),
\end{subeqnarray}
where $\omega_i E_{total}$ is equal to the sum of the above terms. Here, we note that the sum of pressure terms ($P_{up}$+ $P_{vp}$+ $P_{wp}$) and the continuity terms ($P_{pu}+P_{pv}+P_{pw}$) cancel out each other. Also, in the inviscid limit, the Doppler shift terms ($P_{uu}$, $P_{vv}$, $P_{ww}$, $P_{bb}$) are purely imaginary and do not contribute to the growth rate either. Finally, the buoyancy terms $P_{wb}$ and $P_{vb}$ play a role only in exchanging kinetic energy with potential energy in conjunction with the terms $P_{bv}$ and $P_{bw}$ from the density equation. Therefore, the only term that would contribute to the total energy becomes the production by base-flow shear, $P_{uv}$. 

\section{\added{Froude number effect on three-dimensional temporal instability} \label{sec:appen_three}}

\floatsetup[figure]{style=plain,subcapbesideposition=top}
\begin{figure}
    \centering{}
    \sidesubfloat[]{
        \includegraphics[width=0.45\columnwidth]{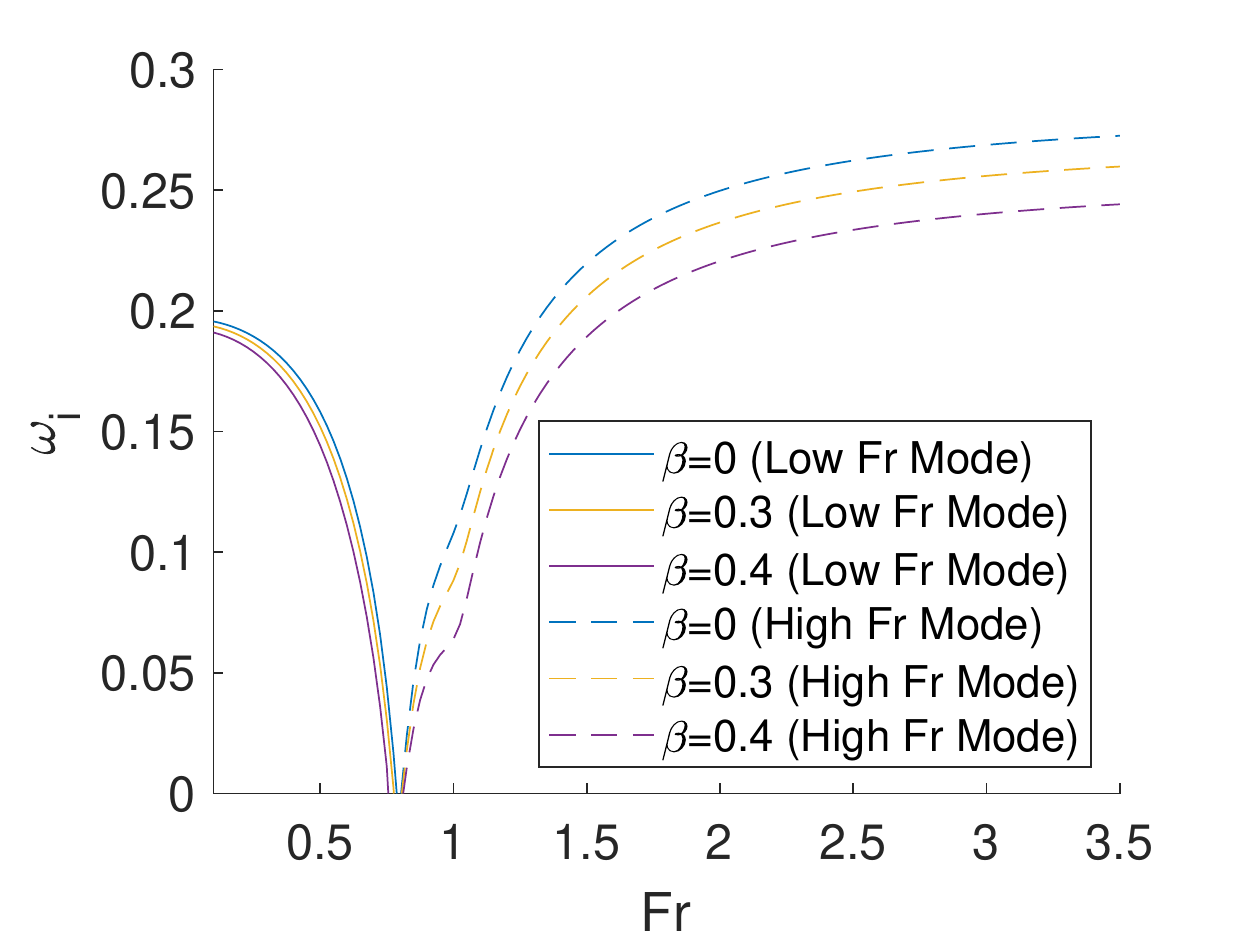}\label{fig:fig6exta}
    }
    \sidesubfloat[]{
        \includegraphics[width=0.45\columnwidth]{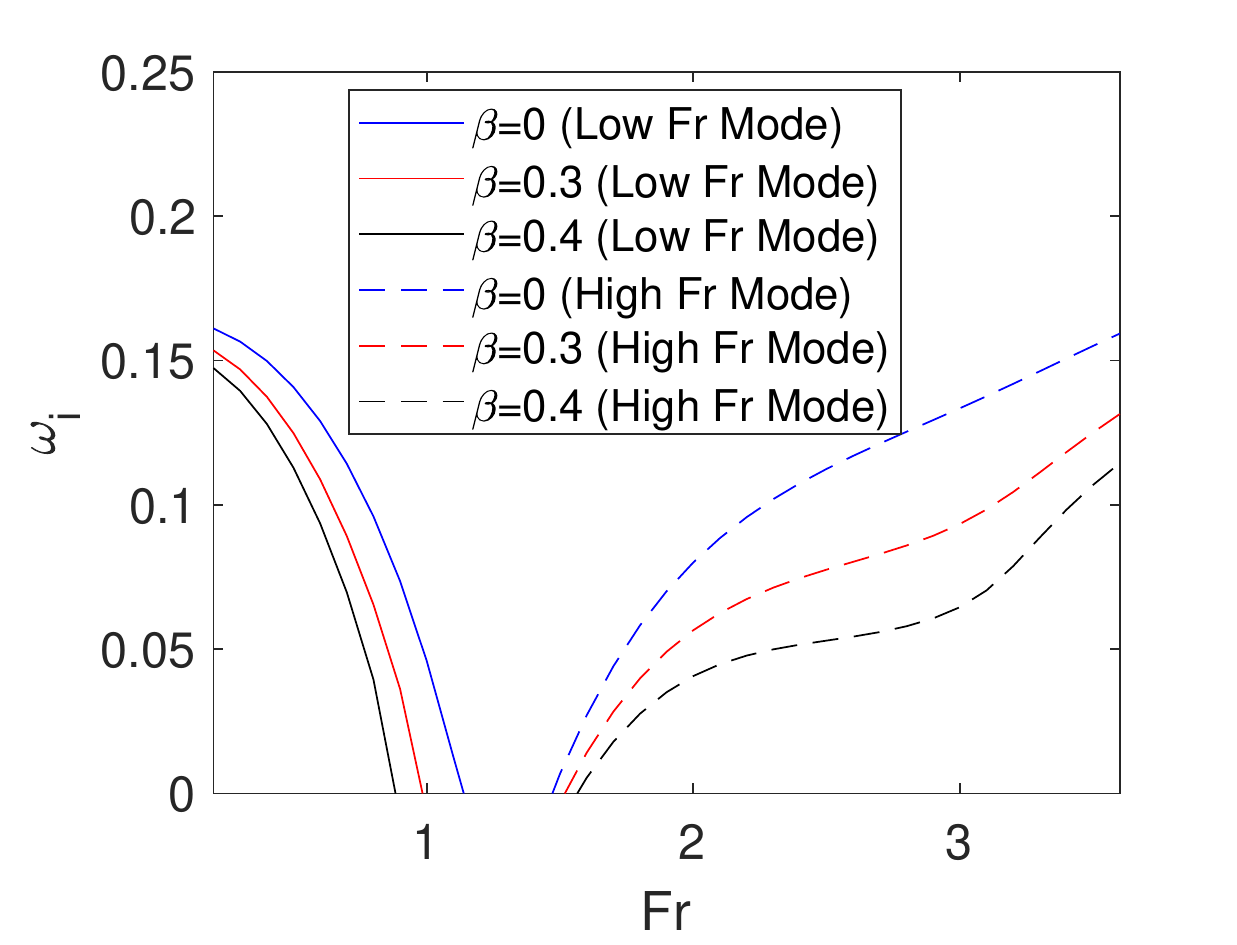}\label{fig:fig6extb}
    }
    \caption{\added{Temporal growth rate of the most unstable mode with respect to Froude number for several $\beta$ at $(a)$ $\theta=30^{\circ}, \Rey=25, \alpha=1$ and $(b)$ $\theta=60^{\circ}, \Rey=50, \alpha=0.4$.} \label{fig:fig6ext}}
\end{figure}

\added{A three-dimensional temporal stability analysis is performed here. Here, we consider several sets of $\alpha \neq 0$ and $\beta \neq 0$ over a range of $\Frou$. The result is shown in figure \ref{fig:fig6ext}. As expected from the analysis in \S\ref{subsec:AU}, the behaviour of three-dimensional instability mode with respect to $\Frou$ is qualitatively the same as that of absolute instability for $\beta=0$.}

\bibliographystyle{jfm}
\bibliography{bibli}

\end{document}